\documentclass[final,10pt,twocolumn,superscriptaddress]{revtex4-2}
\usepackage{amsmath,amssymb,bm}
\usepackage{graphicx,color,multirow}
\usepackage[english]{babel}
\usepackage{float}
\usepackage{microtype}
\definecolor{url}{RGB}{0,20,160}
\usepackage[colorlinks=true,linkcolor=blue,citecolor=blue,urlcolor=url]{hyperref}
\usepackage[usenames,dvipsnames,svgnaes,table]{xcolor}

\usepackage{pdfpages}
\usepackage[T1]{fontenc}
\usepackage{lmodern}

\makeatletter
\AtBeginDocument{\let\LS@rot\@undefined}
\def\frutiger{cmss10 }
\def\frutigerbold{cmssbx10 }
=\frutigerbold at 14pt
=\frutiger at 8pt
=\frutigerbold at 12pt
=\frutigerbold at10pt
=\frutigerbold at 8pt
=\frutiger at 8pt
=\frutigerbold at 31pt
\def\@caption@tabnum@sep{\figtextfont{{ }{\bf\textbar}{ }}}%
\def\fnum@table{{\bf\tablename~\thetable}}
\renewenvironment{table}{\@float{table}\def\textbf##1{{\fignumfont ##1}}\def\bf{\fignumfont}}{\end@float}
\def\@caption@fignum@sep{\figtextfont{{ }{\bf\textbar}{ }}}%
\renewenvironment{figure}{\@float{figure}\def\textbf##1{{\fignumfont ##1}}\def\bf{\fignumfont}}{\end@float}
\renewcommand{\fnum@figure}{\bf Figure \thefigure}
\def\@startsection#1#2#3#4#5#6{%
\if@noskipsec\leavevmode\fi
\par\@tempskipa #4\relax
\@afterindenttrue
\ifdim\@tempskipa <\z@
\@tempskipa -\@tempskipa \@afterindentfalse
\fi\if@nobreak\everypar{}%
\else\addpenalty\@secpenalty\addvspace\@tempskipa\fi
\@ifstar{\@ssect{#3}{#4}{#5}{#6}}{\@dblarg{\@sect{#1}{#2}{#3}{#4}{#5}{#6}}}}
\def\@sect#1#2#3#4#5#6[#7]#8{%
\ifnum #2>0
\let\@svsec\@empty
\else\refstepcounter{#1}\protected@edef\@svsec{\@seccntformat{#1}\relax}\fi
\@tempskipa #5\relax
\ifdim\@tempskipa>\z@
\begingroup#6{\@hangfrom{\hskip #3\relax\@svsec}%
\interlinepenalty \@M #8\@@par}\endgroup
\csname #1mark\endcsname{#7}%
\addcontentsline{toc}{#1}{%
\ifnum #2>\c@secnumdepth\else
\protect\numberline{\csname the#1\endcsname}\fi #7}%
\else\def\@svsechd{#6{\hskip #3\relax
\@svsec #8\ifnum#2=2.\fi}%
\csname #1mark\endcsname{#7}%
\addcontentsline{toc}{#1}{%
\ifnum #2>\c@secnumdepth \else
\protect\numberline{\csname the#1\endcsname}\fi #7}}%
\fi\@xsect{#5}}
\renewcommand\section{\@startsection {section}{1}{\z@}%
{-10pt \@plus -1ex \@minus -.2ex}{.5ex }{\normalfont\Large\bfseries\sectionfont}}
\renewcommand\subsection{\@startsection{subsection}{2}{\z@}%
{10pt\@plus 1ex \@minus .2ex}{-0.5ex \@plus .2ex}{\normalfont\large\bfseries\subsectionfont}}
\def\frontmatter@title@format{\titlefont\centering}%
\def\frontmatter@title@below{\addvspace{-5pt}}%

\newcommand*\bib@heading{%
  \section{\refname}
  \fontsize{8}{10}\selectfont
}
\newcommand*\@openbib@code{%
      \advance\leftmargin\bibindent
      \itemindent -\bibindent
      \listparindent \itemindent
      \parsep \z@
}%
\newdimen\bibindent
\makeatother
\usepackage{gitinfo2}

\usepackage[final]{prelim2e}

\begin{document}

	\title{Discovery of stable surfaces with extreme work functions by high-throughput density functional theory and machine learning}
	\author{Peter Schindler}
	\email{p.schindler@northeastern.edu}
	\affiliation{Department of Mechanical and Industrial Engineering, Northeastern University, Boston, MA 02115, USA}
	\author{Evan R. Antoniuk}
	\affiliation{Materials Science Division, Physical and Life Sciences Directorate, Lawrence Livermore National Laboratory, Livermore, CA 94550, USA}
	\author{Gowoon Cheon}
	\affiliation{Google Research, Mountain View, CA 94043 , USA}
	\author{Yanbing Zhu}
	\affiliation{Department of Applied Physics, Stanford University, Stanford, CA 94305, USA}
	\author{Evan J. Reed}
	\affiliation{Department of Materials Science and Engineering, Stanford University, Stanford, CA 94305, USA}
	\begin{abstract}
The work function is the key surface property that determines how much energy is required for an electron to escape the surface of a material. This property is crucial for thermionic energy conversion, band alignment in heterostructures, and electron emission devices. Here, we present a high-throughput workflow using density functional theory (DFT) to calculate the work function and cleavage energy of 33,631 slabs (58,332 work functions) that we created from 3,716 bulk materials, including up to ternary compounds. The number of materials for which we calculated surface properties surpasses the previously largest database, the Materials Project, by a factor of $\sim$27. On the tail ends of the work function distribution we identify 34 and 56 surfaces with an ultra-low ($<2$ eV) and ultra-high ($>7$ eV) work function, respectively. Further, we discover that the $(100)$-Ba-O surface of BaMoO$_3$ and the $(001)$-F surface of Ag$_2$F have record-low (1.25 eV) and record-high (9.06 eV) steady-state work functions without requiring coatings, respectively. Based on this database we develop a physics-based approach to featurize surfaces and use supervised machine learning to predict the work function. We find that physical choice of features improves prediction performance far more than choice of model. Our random forest model achieves a mean absolute test error of 0.09 eV, which is more than 6 times better than the baseline and comparable to the accuracy of DFT. This surrogate model enables rapid predictions of the work function ($\sim 10^5$ faster than DFT) across a vast chemical space and facilitates the discovery of material surfaces with extreme work functions for energy conversion, electronic applications, and contacts in 2-dimensional devices.
	\end{abstract}
	\maketitle

\section{Introduction}
The work function is a fundamental surface parameter of a material that determines how much energy is required to extract an electron to a field-free region outside the surface; lower work functions facilitate electron emission. Work functions play a key role in technologies that require precise control of contact barriers such as printed and organic electronics.\cite{Zhou2012b,Lindell2006,Dadlani2014} Materials with low work functions are crucial for electron emission devices (THz sources\cite{Snapp2012,Barik2013} and fluorescent light bulbs\cite{Watanabe2011}), electron sources for scientific instruments,\cite{Voss2014,Trenary2012} and high-brightness photocathodes.\cite{Antoniuk2020} Especially for thermionic energy converters (TECs),\cite{Lee2014, Lee2012, Schwede2013} discovery of thermally stable, ultra-low work function materials (less than 1.5 eV) would allow thermionic conversion of heat ($>1500\;^\circ$C) directly to electricity with efficiencies exceeding 30\% (typical thermionic and thermoelectric converters have efficiencies around 10\%). Materials with high work functions play a key role in engineering the electron tunneling barrier in electronics (for example in dynamic RAM applications\cite{Kim2016a} and for contacts in modern 2D-based transistors\cite{Chuang2014}) as well as selective contacts in solar cells\cite{Schulz2016}.\\
The most commonly used materials for low work function applications that are chemically and thermally stable are compounds like lanthanum hexaboride\cite{Pelletier1979,Kanitkar1976} and thoriated tungsten\cite{PhysRev.49.78,PhysRev.53.570,Sillero2010,Bergner2011} with a work function around 2.5 eV. For thermionic converters, extremely low work functions are required, which are achieved by sub-monolayer coatings of alkali metals (most commonly cesium) on metal surfaces. The resulting work functions are much lower than the work function of either metal or coating individually. This effect is due to the partial transfer of electron charge from the adsorbate to the substrate and the resulting formation of surface dipoles that lower the vacuum energy level near the surface.\cite{Chou2012} Perovskite oxides containing transition metals have also shown promise for low work function applications down to 1.6 eV.\cite{Jacobs2016b} Alkaline vanadates (\textit{e.g.}, SrVO$_3$) have also shown promise in thermionic emission devices.\cite{Lin2022Oct} Coatings using cesium or barium combined with oxygen are well known to achieve $\sim 1$ eV work functions in photocathode applications, for instance on III-V semiconductors or silver.\cite{Uebbing1970,James1971,Jacobs2014a} The anti-fluorite X$_2$O family (where X is an alkali metal) are known to exhibit work functions down to 1.2 eV.\cite{Andersson1977,Fecher1994,Jupille1992} Diamondoids\cite{Narasimha2015} and phosphorous-doped diamond thin films have shown similarly low work functions.\cite{Koeck2009} More recently, a work function of 1.01 eV has been achieved by electrostatically gating cesium/oxygen covered graphene,\cite{Yuan2015} which resulted in enhanced TEC efficiency.\cite{Yuan2017} The lowest experimentally measured work function of 0.7 eV has been obtained by inducing a surface photovoltage on Cs/O$_2$ coated Gallium Arsenide.\cite{Schindler2019} The lowest theoretically predicted steady-state value to date is 0.7--0.8 eV for potassium adsorbed on monolayers of MoTe$_2$ or WTe$_2$.\cite{Kim2017}\\
Large work function materials are required for application in hole extraction/injection layers in organic solar cells and light emitting diodes. Thermally evaporated transition metal oxides are commonly used for this purpose, specifically MoO$_3$, WO$_3$, V$_2$O$_5$, NiO, and CrO$_3$, with work functions ranging between 5 and 7 eV.\cite{Greiner2012,Zilberberg2011,Bivour2015} In many cases, organic coatings are used to increase the work function such as poly-styrene sulfonate,\cite{Lim2014,Zhong2020a,Hatton2009}, conducting polyaniline polymers,\cite{Choi2011} fluorinated self assembled monolayers\cite{Fenwick2015}, and benzoic acid trifluoromethyl\cite{Lin2020}. The largest increase in work function was reported for silver incorporating L-cysteine and Zn(OH)$_2$ with a boost from 4.3 to 7.4 eV.\cite{He2020} Transparent conductive oxides such as indium tinoxide and similar derivatives have work functions reported up to 6.1 eV.\cite{Cui2001,Marks2002,Nomoto2020,Helander2011,Zakutayev2011} In metal--insulator--metal devices high work functions are crucial to increase the Schottky barrier-height and hence reduce leakage currents. Ruthenium oxide with a work function of 6.1 eV has been established in these devices.\cite{Schafranek2010} Recently, high work function materials such as MoO$_x$ ($x<3$) and RuCl$_3$ have garnered attention for establishing low resistance contacts in 2D devices, with reported work functions of 6.6 and 6.1 eV, respectively.\cite{Battaglia2014,Guo2014,Wang2020a}\\
In recent years, data-driven approaches based on high-throughput \textit{ab-initio} calculations have emerged as a new paradigm to facilitate the search through vast chemical spaces for new materials with tuned properties or novel behavior. Due to the continued increase in computing power and improvements of theoretical methods, the accuracy of predicted material properties has reached a reliability comparable to experiments while in some cases greatly surpassing them in terms of speed and cost. The rapid increase in available computational data structured in open source material databases such as Materials Project (MP),\cite{Jain2013a} AFLOW,\cite{Curtarolo2012} and NOMAD\cite{Draxl2018} has opened an avenue towards material discovery using data-mining and statistically driven machine learning approaches. However, most big material databases largely lack to report surface properties like the work function as each bulk material typically has dozens of distinct low-index crystalline surfaces and terminations. Each unique surface must be generated and calculated separately increasing the complexity and computational effort required. In the MP database this has only been done for $\sim 100$ polymorphs of elemental crystals.\cite{Tran2016} The largest computational surface catalysis database (>1 million surface adsorption calculations), the OC20, does not report on the work function or surface energy.\cite{Chanussot2021May} Based on several thousand newly predicted 2D materials\cite{Cheon2017} there are two other databases\cite{Haastrup2018,Choudhary2017,Choudhary2023Jan} that report \textit{ab-initio} work functions. It is straight-forward to calculate the work function for 2D materials as they typically have only one relevant surface. The JARVIS-DFT and the C2DB databases have work functions calculated for about $700$ and $4000$ 2D surfaces, respectively. The work function distributions for these 2D material databases are plotted in Figure S1 and the distribution metrics and a list of the 2D materials with the lowest and highest work functions is compiled in Table S1.\\
Some statistical analyses have been carried out in literature showing that the electronegativity is linearly correlated with the work function both for elemental crystals and binary compounds.\cite{Tran2016, Yamamoto1974} Additionally, for elemental crystals an inverse correlation with the atomic radius is pointed out. The work function of elemental crystals ranges between 2 and 6 eV (for Cesium and Selenium, respectively). The statistical analyses of about 30 binary compounds shows that a correlation between the electronegativity of the atom with the lower electronegativity is the strongest (better than arithmetic or geometric mean of the individual electronegativities). Density functional theory has been a well-established approach (using a slab configuration) to calculate the work function, similar to the more simplistic Jellium model.\cite{Lang1971} Also a phenemonolgical model has been developed that is able to estimate the work function fairly accurately for metals and alkaline-metal coated surfaces.\cite{Brodie2014} This phenomenological equation is a function of the atomic radius and the number of atomic sites per unit cell area. However, it relies on a single parameter (loosely related to the number of electrons that an atom can donate to the surface) that is not clearly defined for more complex surfaces and takes on nonphysical values in the case of alkaline coatings. In recent work, Hashimoto et al.\cite{Hashimoto2020a} attempted to screen for low and high work function materials using a Bayesian optimization approach. However, they assume the work function to be approximated solely as a bulk property neglecting any surface contributions during screening. For the highest and lowest ``bulk work function" material candidates the actual surface contributions have then been included which rendered most of their top candidate materials to exhibit average work functions between 3 and 6 eV. Unsurprisingly, among their top candidate materials, they have found that the (110) surface of elemental Cesium has a low work function of 2.0 eV and that the (111) surface of KEuO$_2$ has a relatively high work function of 8.4 eV. The approximated bulk work function of some of the screened work function candidates differs as much as 7 eV from the actual work function when including the surface contributions. This clearly shows that, while for simple structures (such as elemental metals) the work function can theoretically be predicted from bulk properties alone,\cite{Halas2010} it is important to consider surface contributions to quantitatively predict the work function of a material. The surface termination, atom adsorption (most commonly oxygen and hydrogen), contamination, and reconstructions can affect the surface dipole and hence the effective work function. While a crystal graph convolutional neural network has been used successfully to predict the cleavage energies of intermetallic slabs,\cite{Palizhati2019Nov} there has been no reports on featurizing slabs to predict the work function (except for the MXene 2D-material class\cite{Roy2023May}).\\
In this paper, we use high-throughput density functional theory (DFT) to calculate 58,332 surface work functions and 33,631 cleavage energies based on 3,716 bulk crystal structures (up to ternary compounds with a band gap less than 0.1 eV). The created database gives insight into work function trends observed across a large chemical space. Based on the database we develop a machine learning model with a low mean absolute test-error of 0.09 eV which is more than 6 times lower than the baseline (\textit{i.e.}, predicting every surface to have the database-average work function) and about 4 to 5 times better than state-of-the-art benchmarking machine learning models (automatminer and Coulomb matrix). The database and machine learning model enable us to identify several promising, stable material surfaces with extremely low ($<2$ eV) and extremely high ($>7$ eV) work functions. Further, the work function model established in this work enabled the discovery of new ultra-bright photocathode materials reported in  \cite{Antoniuk2021Nov,Antoniuk2020Jun}.

\section{Methods}

\begin{figure*}[tp]
\includegraphics[width=\textwidth]{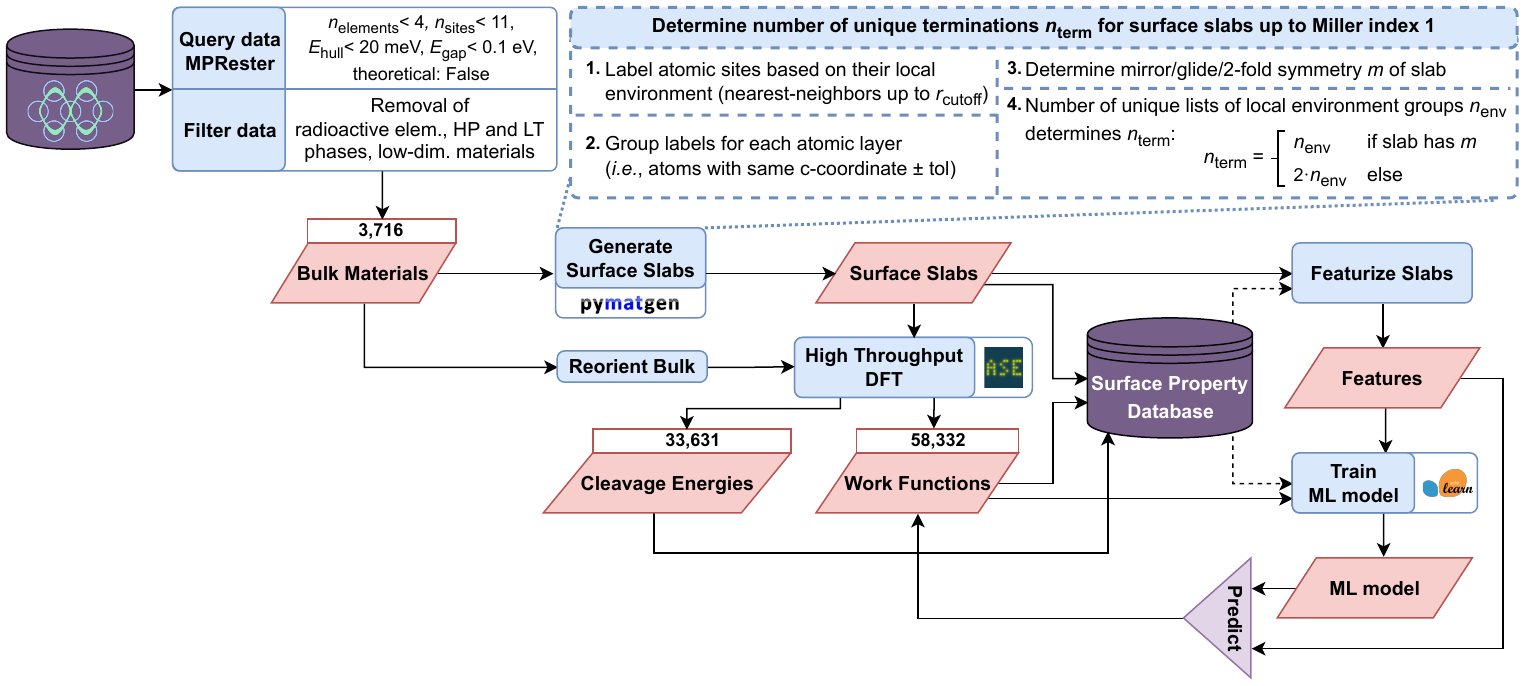}
\caption{Workflow of the creation of the surface property database and surrogate machine learning model. The illustration includes the steps for material selection, high-throughput DFT calculations, surface slab creation, and supervised machine learning predictions. The dashed block details the procedure of determining the unique terminations of all surfaces up to a Miller index of 1.
}%
\label{fig:workflow}
\end{figure*}

\subsection{Materials Selection and High-throughput Workflow}
The workflow of the work function database's creation is illustrated in Figure \ref{fig:workflow}. A total of 3,716 crystal structures were queried from the Materials Project (on 2/2/2023) using the REST framework.\cite{Jain2013a,Ong2015} Up to ternary materials with 10 or less atoms in the conventional unit cell were considered that have an energy above hull of less than 20 meV/atom, are metallic ($E_\mathrm{gap}<0.1$ eV), and are tagged as an experimental structure (\textit{i.e.}, there exists at least one ICSD entry that reports the corresponding material as experimentally synthesized). Materials with an element present that is radioactive, a noble gas or from the actinide group were excluded. Further, materials that have experimental tags that indicate high pressure or low temperature conditions, as well as low-dimensional materials were excluded as well. The frequency with which each chemical species appears in the database (bulk compounds) is plotted as a heat-map on the periodic table in Figure S3.\\
From this set of materials, surfaces up to a Miller index of 1 were generated using Pymatgen’s surface module.\cite{Ong2013,Tran2016,Tran2019} Each surface Miller index generally has more than one unique surface termination. The Pymatgen surface module has a built-in option to generate slabs with different terminations determined by possible shifts in the $c$-direction. We have developed an alternate algorithm to ensure that we generate all possible \textit{unique} terminations based on the local environment of surface atoms, as summarized in the dashed block in Figure \ref{fig:workflow} and described in more detail in the Supplementary Material. \\
According to $n_\mathrm{term}$ we generate slabs that contain the appropriate number of atomic layers to expose each unique termination. Further, we minimize the number of slabs required for the DFT calculations by having two distinct terminations on either side of the slab, whenever possible. The initial slab thickness is minimized while still ensuring that after all necessary subtractions the final slab is at least 10 $\AA{}$ thick. Following this procedure, we create 37,163 slabs and 17,998 bulk structures of which 36,962 and 17,993 converged respectively during self-consistent field calculations with a total computational time of around 440,000 core-hours. The converged calculations after duplicate removal (described in the machine learning model training section) returned a total of 58,332 work functions and 33,631 cleavage energies of which computational details are described next.

\subsection{First Principles Calculations}
The total energies of the slab and bulk as well as the work functions are calculated by gradient-corrected DFT using the PBE exchange-correlation functional.\cite{PBE} Self-consistent, periodic, total energy calculations are performed using Quantum Espresso (v.7.1).\cite{QE} Ultrasoft Vanderbilt pseudopotentials\cite{Garrity2014} are used to describe core electron interactions ($f$-electrons not included for lanthanides\cite{Corso2014}) and the Kohn-Sham one-electron valence states are expanded in a plane wave basis set with a kinetic energy cutoff of 550 eV. The electron density is expanded up to ten times the energy of the plane wave energy cutoff. An extra 30 unoccupied bands are added for holding valence electrons to improve convergence. All slabs generated by the high-throughput procedure described above have a minimum thickness of 10 $\AA{}$ and 15 $\AA{}$ of vacuum between periodic slab repetitions in the $c$-direction to preclude interactions between periodic images. Brillouin zone sampling is performed under a grid spacing of less than $0.05 \;\mathrm{\AA{}}^{-1}$ with finite-temperature Gaussian smearing ($\sigma = 0.1$ eV). A dipole correction is applied in the $c$-direction. The work function, $\phi$, is determined by the difference of the electrostatic energy in the vacuum region, $E_\mathrm{vac}$, and the Fermi energy, $E_F$: $\phi=E_\mathrm{vac}-E_F$. The PBE exchange-correlation functional has previously been shown to give reliable work functions for elemental crystals in agreement with experimental values with a slight systematic underestimation. The errors with respect to experiment have been reported to be below 0.3 eV for elemental crystals\cite{DeWaele2016} and $\sim 0.15$ eV for elemental metals.\cite{Patra2017a} It is also worth noting that work functions measured experimentally can significantly vary based on the technique and specimen (single- vs. poly-crystalline) used.\cite{Lin2023Mar} The DFT calculation inputs for Quantum Espresso are automatically generated with the atomic simulation environment (ASE)\cite{Hjorth_Larsen_2017} Python package and submitted into a high performance computing queuing system (SLURM) using job arrays.\\
To estimate the convergence accuracy of the DFT-calculated work functions we reran $\sim 1\%$ of the database (randomly selected) with stricter convergence parameters (energy cutoff of 850 eV and Brillouin zone sampling with a grid spacing of $\leq 0.02 \;\mathrm{\AA{}}^{-1}$). The resulting mean absolute error and root-mean square error of the work function are 22 and 30 meV, respectively.\\
The cleavage energies have been calculated via total energy calculations of the slab, $E_\mathrm{slab}$, and the respective bulk, $E_\mathrm{bulk}$, with the same unit cell orientation as the slab. This reorientation ensures more efficient convergence with slab thickness due to $k$-point matching of the two calculations.\cite{Sun2013a} The cleavage energy is defined as
\begin{equation}
E_\mathrm{cleavage} = \frac{E_\mathrm{slab}-n_\mathrm{bulk}\cdot E_\mathrm{bulk}}{2\cdot A_\mathrm{slab}}=\frac{\gamma_\mathrm{top}+\gamma_\mathrm{bottom}}{2}
\end{equation}
where $n_\mathrm{bulk}$ is the number of bulk unit cells contained in the slab and $A_\mathrm{slab}$ is the surface area of the slab. The surface energies are denoted as $\gamma_\mathrm{top}$ and $\gamma_\mathrm{bottom}$ for the top and bottom surface of the slab, respectively. Using this definition, the cleavage energy is equal to the surface energy, $E_\mathrm{cleavage}=\gamma$, for the case of symmetric slabs (\textit{i.e.}, there exists a mirror or glide plane parallel to the surface, or a 2-fold rotation axis normal to the surface). For non-symmetric slabs, the upper bound for both surface energies is $2\cdot E_\mathrm{cleavage}>\gamma_\mathrm{top,bottom}$. This workflow has also been used to calculate adhesive properties in electrolyte interfaces for solid-state battery applications.\cite{Ransom2023Sep}\\
For the slabs that exhibit extreme work functions, ionic relaxation calculations of the topmost surface atoms (atoms within 3$\AA{}$ of the surface were relaxed, the remaining atoms fixed) of the $1\times 1$ slabs were converged to a force less than 0.05 eV$/\AA{}$. Only the surface of interest was ionically relaxed while the opposite side remained fixed.
\subsection{Machine Learning Model Training}
The dataset is randomly split into training and test sets (90/10 split) and the hyperparameters are optimized with a grid-search implementing 10-fold cross-validation on the training set. Multivariate linear regression, random forest, and neural network models are set up with the scikit-learn package in Python. The custom featurization procedure is laid out in the results and discussion section. Surfaces that had identical features to any other surface in the dataset (and a difference in work function between the two surfaces of $<0.1$ eV) were removed as duplicates from the dataset before training. For benchmarking purposes we used the automatminer testing suite\cite{Dunn2020} (200 features) and a conventional Coulomb matrix (topmost 5 surface atoms as input, matrix sorted by $\ell_2$-norm of columns -- the flattened matrix yields 25 features that are used in a random forest model).\cite{Himanen2020} For automatminer we use the ``express" setting and for comparison we used the bulk unit cell and the topmost 5 atomic layers of the surface slabs as inputs. As a baseline model we predict the work function to be the average work function regardless of the surface.

\section{Results and Discussion}

\begin{figure*}[tp]
\includegraphics[width=0.95\textwidth]{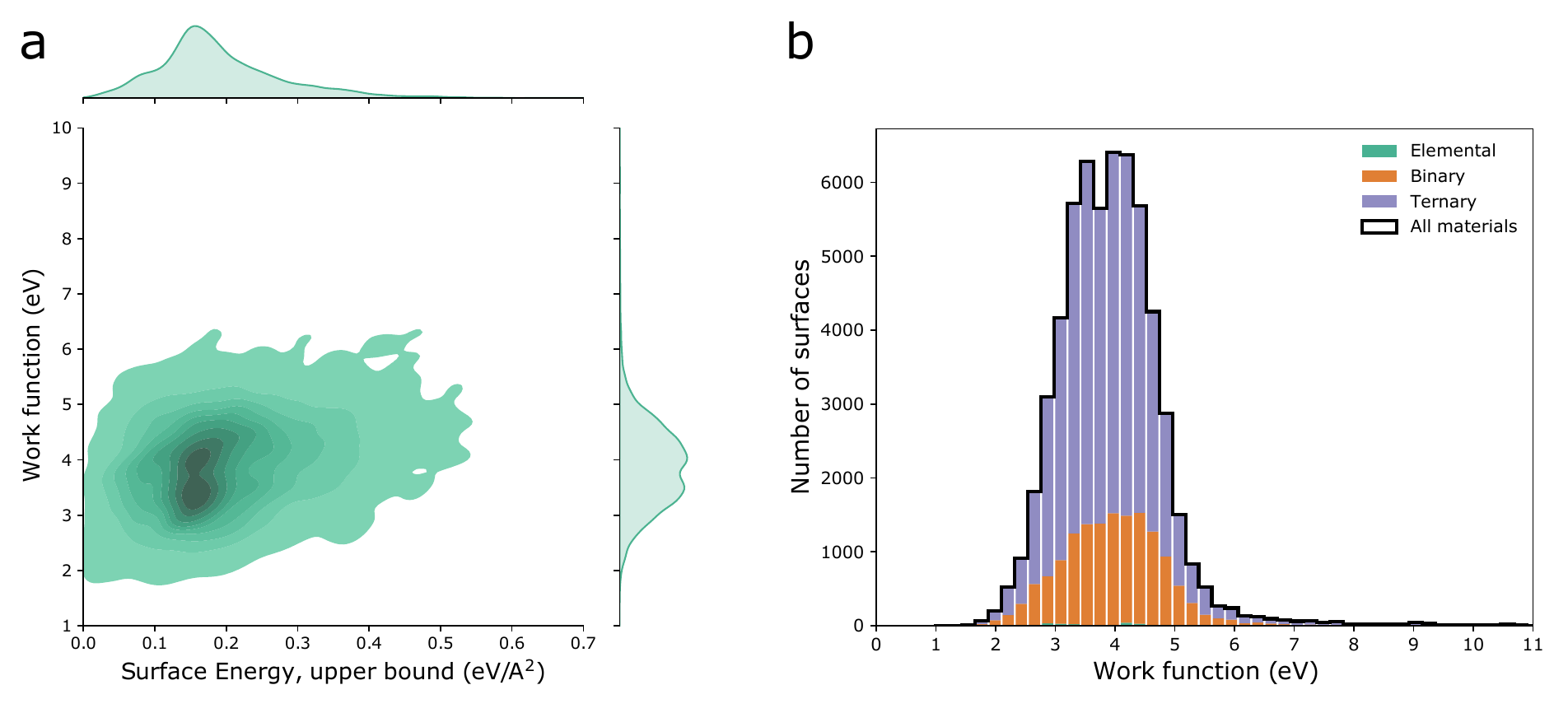}
\caption{\textbf{a} Kernel density estimate plot illustrating the distribution of the work function vs. the upper bound of the surface energy. Both calculated by DFT for unrelaxed slabs. \textbf{b} Distribution of the work functions plotted as a stacked histogram. Outline corresponds to overall distribution under which the stacked, colored bars signify the number of surfaces based on elemental, binary, and ternary compounds. The average of the distribution is 3.92 eV with a standard deviation of 0.86 eV.
}%
\label{fig:WF_distr}
\end{figure*}

\begin{table}
\begin{tabular}{cccccc}
\hline
& \textbf{Number} & \textbf{Average}& \textbf{St.Dev.}& \textbf{Min.} & \textbf{Max.}\\ 
\hline
Elemental & 261 & 3.72 & 0.84 & 2.30 & 5.70\\
Binary & 14,623 & 4.00 & 0.91 & 1.43 & 10.54\\
Ternary & 43,448 & 3.89 & 0.84 & 1.17 & 11.07\\
\hline
All & 58,332 & 3.92 & 0.86 & 1.17 & 11.07\\
\hline
\end{tabular}
\caption{Detailed work function distribution metrics for elemental, binary, and ternary compounds for unrelaxed slabs. All values in eV.} 
\label{tab:distr}
\end{table}

\begin{table}

\begin{tabular}{ccccccc}
\hline
&\textbf{Sym.}&\textbf{Number}&\textbf{Average}& \textbf{St.Dev.}& \textbf{Min.} & \textbf{Max.}\\ 
\hline
Elemental & Yes & 206 & 81.5 & 57.0 & 5.4 & 240.7\\
Elemental & No & 30 & 42.4 & 25.1 & 5.7 & 87.8\\
Binary & Yes & 3,596 & 88.7 & 50.2 & 5.9 & 396.3\\
Binary & No & 5,819 & 109.0 & 58.1 & 7.1 & 397.3\\
Ternary & Yes & 2,229 & 87.3 & 37.0 & 1.0 & 296.4\\
Ternary & No & 21,751 & 102.0 & 41.8 & 0.3 & 388.5\\
\hline
All & Yes & 6,031 & 88.0 & 46.0 & 1.0 & 396.3\\
All & No & 27,600 & 103.4 & 45.8 & 0.3 & 397.3\\
\hline
\end{tabular}
\caption{Detailed cleavage energy distribution metrics for elemental, binary, and ternary compounds for unrelaxed symmetric and asymmetric slabs. All values in meV/$\AA{}^2$.} 
\label{tab:distr_CE}
\end{table}

\subsection{Analysis of Surface Database}
\begin{figure*}[tp]
\includegraphics[width=0.95\textwidth]{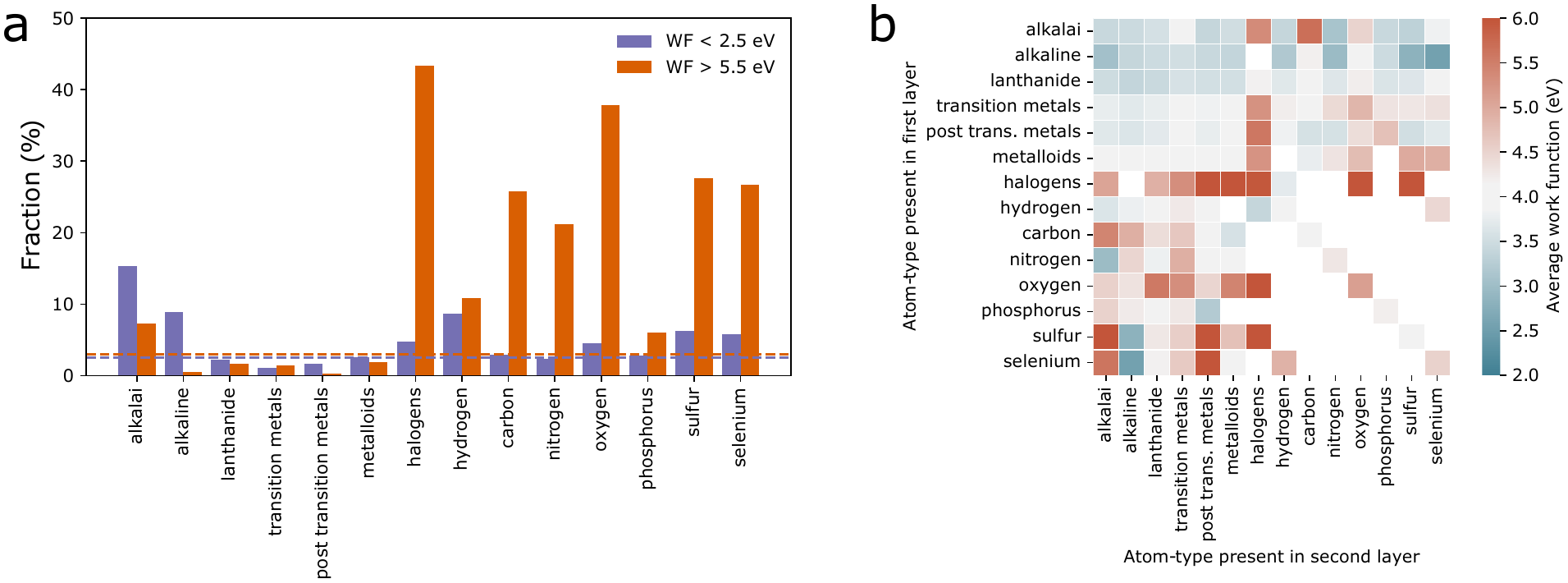}
\caption{Work function trends observed in the database. \textbf{a} Fraction of material surfaces that have a work function below 2.5 eV (purple) and above 5.5 eV (orange) is shown depending on which type of chemical species is present at the topmost surface. The dashed lines indicate the average fraction across the entire database regardless of chemical species at the surface. \textbf{b} Heat-map of the average work function plotted as a function of chemical species present at the topmost layer (vertical axis) and second atomic layer (horizontal axis). The color bar displays work functions below and above average as blue and red, respectively. Categories with a population of less than 10 surfaces have been left blank.
}%
\label{fig:WF_trends}
\end{figure*}

First, we analyze the surface property database created by high-throughput DFT (33,631 slabs based on 3,716 bulk crystals) in terms of its distribution and trends in the studied chemical space. A contour plot of the database distribution of DFT-calculated work functions vs.\ surface energies is shown in Figure \ref{fig:WF_distr}a for ionically unrelaxed slabs. The calculated cleavage energy is exactly equal to the surface energy in the case of symmetric slabs that exhibit either a mirror or glide plane parallel to the surface, or a 2-fold rotation axis normal to the surface. For slabs lacking this symmetry, $2\cdot E_\mathrm{cleavage}$ is used as an upper bound to the surface energy. The work function distribution of ionically unrelaxed surfaces is plotted in Figure \ref{fig:WF_distr}b and shows a near-Gaussian distribution with an extended tail towards higher work functions. The average of the entire distribution is at 3.92 eV with a standard deviation of 0.86 eV, ranging from a minimum to a maximum work function of 1.17 to 11.07 eV, respectively. The stacked bar-chart signifies which proportion of the surfaces within each bin stems from an elemental, binary, or ternary compound. Their distribution metrics are given in Table \ref{tab:distr}. The distribution of cleavage energies for elemental, binary, or ternary compounds are given in Table \ref{tab:distr_CE} for both symmetric and asymmetric slabs, and plotted as a stacked bar histogram in Figure S11. The average of the cleavage energy for all symmetric/asymmetric slabs is 88.0/103.4 meV$/\AA{}^2$ with a standard deviation of 46.0/45.8 meV$/\AA{}^2$, ranging from a minimum to a maximum cleavage energy of 1.0/0.3 to 396.3/397.3 meV$/\AA{}^2$, respectively.\\
The observation that the distribution in work functions is near-Gaussian could indicate that the chemical space we chose was diverse enough to evenly sample work functions across possible values. The extended tail at the high work function end appears to be an artifact coming from ionically unrelaxed surfaces where a small, electronegative atom (\textit{e.g.}, oxygen or hydrogen) is cleaved at a large, unphysical distance (as discussed in the next section and corroborated by Figure S10). This might also be the case for the low work function tail but appears to be less pronounced. This artifact can be mitigated by ionically relaxing the surface slabs (see next section) and we expect this to result in an overall slightly narrower distribution. Interestingly, the work function distributions of binary and ternary compounds (and to a certain extent also the elemental crystals) have similar averages, standard deviations, and ranges. This may be explained by the observation that the work function is primarily determined by the chemical species present in the topmost layer at the surface (as discussed in the next paragraph), and will largely not depend on the total number of chemical species present in the entire unit cell. Moreover, the average work function of the database is lower than the average work function for the JARVIS database and C2DB (4.91 and 5.43 eV, respectively, \textit{cf.} Table S1) while the standard deviations are somewhat similar (1.22 and 1.08 eV, respectively). The average cleavage energy of all asymmetric slabs (103.4 meV$/\AA{}^2$) is higher than the average for all symmetric slabs (88.0 meV$/\AA{}^2$). This is expected because this database is calculated for unrelaxed slabs and cleaving asymmetric slabs may lead to dangling atoms in nonphysical positions too far/close to the other surface atoms. \\
Trends in the work function based on which chemical species are present at the surface are shown in Figure \ref{fig:WF_trends}. The fraction of surfaces with a low work function ($<2.5$ eV, \textit{i.e.}, roughly 1.5 times the standard deviation below average) is especially high for surfaces with alkali or alkaline metals present in the topmost atomic layer. Conversely, the fraction of surfaces with a high work function ($>5.5$ eV, \textit{i.e.}, roughly 1.5 times the standard deviation above average) is especially high for surfaces with halogens, carbon, nitrogen, sulfur, selenium or oxygen present in the topmost atomic layer (\textit{cf.} Figure \ref{fig:WF_trends}a). Surfaces with hydrogen present exhibit more of both low and high work functions than average, likely due to complex chemistries. The total number of surfaces (rather than fractions) are shown in Figure S4. Overall, 48.8\% (29.5 \%) of surfaces that exhibit a work function below 2.0 (2.5) eV contain either alkali or alkaline metals in the topmost atomic layer. Conversely, 54.1 \% (38.0 \%) of surfaces that exhibit a work function above 6.0 (5.5) eV contain either carbon, oxygen, or halogens in the topmost atomic layer.\\
The average work function is plotted as a heat-map based on the chemical species present in the topmost two atomic layers. The trends observed in Figure \ref{fig:WF_trends}a are also seen in the average work function trend in Figure \ref{fig:WF_trends}b. However, one can also observe trends based on combinations of chemical species in the topmost and second atomic layers. For example, the work function average is high for surfaces where halogens are present in the first or second layer. In contrast, the work function average is low for surfaces with alkali or alkaline metals present in the first layer and sulfur or selenium present in second layer -- however, the work function average is high for the reversed layers (\textit{i.e.}, sulfur/selenium in topmost layer). Further trends are plotted in Figures S5 (barchart of average work function as a function of the chemical species present at the topmost layer) and S6 (heat-maps showing percentage and total number of low and high work function surfaces as a function of chemical species present in the top two layers).\\
The trends described generally agree with the chemical intuition that surfaces terminated with electropositive atoms from the alkali or alkaline groups give a low work function, whereas electronegative atoms from the non-metal groups cause increased work functions. However, it is worth noting that while $\sim 15\%$ of surfaces that have an alkali/alkaline metal present in the topmost atomic layer have work function below 2.5 eV, still $\sim 85\%$ have work functions above 2.5 eV (of which $\sim 7 \%$ even have a work function above 5.5 eV), contrary to chemical intuition. This shows that for complex chemistries (\textit{e.g.}, hydrogen-containing compounds) basic chemical intuition is insufficient to predict the work function.

\subsection{Surfaces with Extreme Work Functions}
Material surfaces found on the tail ends of the work function distribution in Figure \ref{fig:WF_distr} are of interest for energy conversion and electronic device applications. However, the calculated \textit{ab-initio} work functions are based on surfaces without considering ionic relaxations. Hence, to screen for viable ultra-low and ultra-high work functions we carry out ionic relaxation calculations for unrelaxed slabs that have a cleavage energy less than 200 meV/$\AA{}^2$ and either surface with a work function that is lower than 2 eV or higher than 7 eV. From 340 slabs on the extreme ends of the distribution, 284 slabs converged during ionic relaxation to a force less than 0.05 eV$/\AA{}$. Atoms within the top 3$\AA{}$ of the topmost atom were allowed to relax while the coordinates of the remaining atoms in the slab were fixed. The cleavage energy is generally reduced through ionic relaxation (\textit{cf.} Fig. S12). As expected, the work functions for materials on the low-end shift towards larger values (mean signed deviation $= +0.20$ eV). The work functions for materials on the high-end shift towards lower values (mean signed deviation $= -0.62$ eV). Hence, this effect is much more pronounced for the high work function tail-end, as can be observed in Figure S10. This effect is also more pronounced for non-symmetric slabs compared to symmetric ones. However, even after ionic relaxation there are many material candidates that retain record-low and -high work functions.\\
Ionically relaxed surfaces with the lowest ($<2$ eV) and highest work functions ($>7$ eV) are listed in Table \ref{tab:extremeWF}. Surfaces created from structurally equivalent materials (but with distinct atom-type occupations) are grouped into material families using a notation of chemical formulas with placeholders accompanied by a list of possible elements present for each placeholder. The compound that resulted in the lowest/highest work function is listed alongside its work function value and cleavage energy for each family. The Miller indices and the atom termination of the surfaces yielding extreme work functions are specified. Polar surfaces have been excluded and the symmetry of the non-polar slabs are indicated in the rightmost column. A slab which indicates "n" (no) in the "Sym." (symmetry) column refers to slabs that require off-stoichiometric slabs to ensure that the top and bottom surfaces are equivalent. All listed slabs are non-polar regardless of the symmetry tag (\textit{i.e.}, for each listed candidate with the "n" tag there exists an off-stoichiometric slab that is symmetric). For slabs that state "y" (yes) in the symmetry column, the cleavage energy is equal to the surface energy, otherwise $2\cdot E_\mathrm{cleavage}$ is an upper bound to the surface energy.\\
Among the discovered low work function candidate surfaces the most common terminations are Ba, Sr, and Cs, in agreement with chemical intuition and previous literature reports (\textit{e.g.}, cesium oxide). The lowest work function after ionic relaxation was calculated for the (100)-Ba-O termination of BaMoO$_3$ at 1.25 eV with a cleavage energy of 67.2 meV/$\AA{}^2$. To our knowledge, this is the lowest steady-state work function computationally predicted in literature for a metallic surface without requiring any (alkaline) monolayer-coating. Further, we identify the (100)-Cs-Cl surface of CsScCl$_3$ to exhibit an ultra-low work function of 1.42 eV in addition to an extremely low surface energy of 5.5 meV/$\AA{}^2$, making it the most promising low work function surface with high stability. Similarly, the (110)-Sr surface of SrN$_2$ and (010)-Ba-Si surface of BaSi are at the Pareto-front of low work function/high stability.\\
The ionically relaxed surfaces exhibiting high work functions  tend to contain fluorine or oxygen in the topmost atomic layer. Similar to the low work function case, this also agrees with chemical intuition as the direction of the surface dipole moment depends on whether the terminating atom is electropositive or electronegative, resulting in a lower or higher vacuum energy level, respectively. The highest work function after ionic relaxation was calculated for the (001)-F surface of Ag$_2$F at 9.06 eV with a cleavage energy of 64.2 meV/$\AA{}^2$. To our knowledge, this is the highest steady-state work function of a metallic surface computationally predicted in literature. It is interesting to note that work functions exceeding 7-8 eV are rarely reported in literature. We further reveal that the (011)-F surface of CuAuF$_5$ and the (10$\bar{1}$)-F surface of CuSnF$_6$ both have extremely high work functions (7.78 and 8.00 eV) while having low surface energies (11.4 and 32.3 meV/$\AA{}^2$).\\
Overall, we have identified 34 distinct ultra-low work function, and 56 high work function surfaces. We note that in a previous version of the work function database, in which we included materials with non-zero band gaps, we further identified low work function candidate materials in agreement with previous literature reports, such as alkaline vanadates (\textit{e.g.}, BaVO$_3$) and alkali-based anti-fluorite structures (\textit{e.g.}, Rb$_2$O, Rb$_2$S). For a discussion on the previous version of the database (and why we excluded it) refer to the supplementary materials. A full list of low and high work function candidate surfaces identified in the previous version of the database are given in Tables S2 and S3, respectively. From the curated lists of discovered surfaces, further analyses can be carried out in future research to assess viability for device applications, such as assessing stability in air/moisture and surface reconstructions.
\begin{table*}
\setlength\extrarowheight{1pt}
\footnotesize
\begin{tabular}{lllllllll}
 & \textbf{sg}  & \multicolumn{2}{l}{\textbf{Material Family}}    & \parbox{2.0cm}{\raggedright\textbf{Comp. with}\\[0\baselineskip] \textbf{min/max} \boldmath$\phi$}  & \textbf{Miller planes - Termination}  & \boldmath$\phi\;(\mathrm{eV})$  & \parbox{1.5cm}{\raggedright \boldmath$E_\mathrm{cleavage}$\\[0\baselineskip]\boldmath$(\mathrm{meV}/\mathrm{\AA{}}^2)$} & \textbf{Sym.} \\ \hline\hline
\multirow{12}{0.3cm}{\rotatebox[origin=c]{90}{\centering \textbf{Low $\phi$ surfaces}}} & 2   &         &                              & CsHg                  & (11$\bar{1}$)-Hg                    & 1.79 & 8.9             & y         \\ \cline{2-9}
 & 63  & Ba\textcolor{purple}{A}     & \textcolor{purple}{A}=Sn,Si                      & BaSi                  & (110)-Ba-\textcolor{purple}{A}, (010)-Ba-\textcolor{purple}{A}       & 1.69 & 16.0            & y         \\ \cline{2-9}
 & 123 & BaMg$_4$\textcolor{purple}{A}$_3$ & \textcolor{purple}{A}=Si,Ge                      & BaMg$_4$Si$_3$              & (001)-Ba                     & 1.71 & 72.0            & n         \\ \cline{2-9}
 &     &         &                              & SrFeO$_2$                & (100)-Sr                     & 1.58 & 67.0            & n         \\ \cline{2-9}
 & 129 & BaMg\textcolor{purple}{A}   & \textcolor{purple}{A}=Ge,Si                      & BaMgSi                & (001)-Ba                     & 1.65 & 15.7            & y         \\ \cline{2-9}
 & 139 & Ba\textcolor{purple}{A}$_2$\textcolor{blue}{B}$_2$  & \parbox{2.5cm}{\vspace{.2\baselineskip}\raggedright \textcolor{purple}{A}=Ag,Zn,Cu,Pd,\\[0\baselineskip] \textcolor{blue}{B}=Ge,Si,As,Sb} & BaPd$_2$Sb$_2$              & (101)-Ba, (101)-Ag, (001)-Ba & 1.73 & 51.9            & n         \\ \cline{2-9}
 &     &         &                              & SrN$_2$                  & (110)-Sr                     & 1.59 & 22.6            & y         \\ \cline{2-9}
 & 164 &         &                              & BaSi$_2$                 & (100)-Ba, (100)-Si           & 1.68 & 45.4            & n         \\ \cline{2-9}
 & 193 &         &                              & Cs$_3$O                  & (110)-Cs, (101)-Cs           & 1.76 & 8.1             & n         \\ \cline{2-9}
 & 194 &         &                              & CsScCl$_3$               & (100)-Sc, (100)-Cs-Cl        & 1.42 & 5.5             & y         \\ \cline{2-9}
 & 221 & Ba\textcolor{purple}{A}O$_3$   & \textcolor{purple}{A}=Nb,Mo                      & BaMoO$_3$                & (100)-Ba-O                   & 1.25 & 67.2            & n \\ \hline\hline


\multirow{12}{0.3cm}{\rotatebox[origin=c]{90}{\centering \textbf{High $\phi$ surfaces}}} & 2   &       &          & CuAuF$_5$  & \parbox{4.5cm}{\vspace{.2\baselineskip}\raggedright (011)-F, (011)-Cu-F, (100)-Au,\\[0\baselineskip](101)-F, (10$\bar{1}$)-F, (110)-Au,(1$\bar{1}$0)-F,\\[0\baselineskip](1$\bar{1}$0)-Au,(111)-F, (11$\bar{1}$)-F,(1$\bar{1}$1)-F} & 7.78 & 11.4 & y \\ \cline{2-9}
 &     & Cu\textcolor{purple}{A}F$_6$ & \textcolor{purple}{A}=Mo, Sn & CuSnF$_6$  & \parbox{5.9cm}{\vspace{.2\baselineskip}\raggedright(10$\bar{1}$)-F, (101)-Cu, (110)-F, (11$\bar{1}$)-F,\\[0\baselineskip](01$\bar{1}$)-Cu, (100)-F, (1$\bar{1}$0)-Cu}                                     & 8.00 & 32.3 & y \\ \cline{2-9}
 & 12  &       &          & CuAgO$_2$  & (001)-O, (01-1)-O                                                                                  & 8.19 & 81.4 & n \\ \cline{2-9}
 & 14  & \textcolor{purple}{A}F$_2$   & \textcolor{purple}{A}=Cu,Ag  & AgF$_2$    & \parbox{4.5cm}{\vspace{.2\baselineskip}\raggedright(01-1)-\textcolor{purple}{A}, (011)-\textcolor{purple}{A}, (01-1)-F,(001)-F, (011)-\textcolor{purple}{A},\\[0\baselineskip] (100)-F, (101)-F, (110)-F, (11-1)-F}                         & 7.01 & 22.2 & y \\ \cline{2-9}
 & 15  &       &          & AgO     & (010)-O                                                                                            & 7.45 & 68.7 & n \\ \cline{2-9}
 & 127 &       &          & KCuF$_3$   & (110)-Cu-F                                                                                         & 8.55 & 49.8 & n \\ \cline{2-9}
 & 136 &       &          & PbO$_2$    & (101)-O-Pb, (101)-O, (111)-Pb-O                                                                      & 7.20 & 61.4 & y \\ \cline{2-9}
 & 162 &       &          & As$_2$PdO$_6$ & (001)-O                                                                                            & 8.13 & 48.9 & n \\ \cline{2-9}
 &     &       &          & NbHg$_3$F$_6$ & (001)-F                                                                                            & 8.27 & 22.7 & n \\ \cline{2-9}
 & 164 &       &          & Ag$_2$F    & (001)-F                                                                                            & 9.06 & 64.2 & n \\ \cline{2-9}
 &     &       &          & Rb$_2$RhF$_6$ & (110)-F                                                                                            & 7.65 & 68.8 & n \\ \cline{2-9}
 & 176 & \textcolor{purple}{A}Cl$_3$  & \textcolor{purple}{A}=Eu,Ce  & EuCl$_3$   & (100)-\textcolor{purple}{A}-Cl, (101)-Cl, (110)-\textcolor{purple}{A}-Cl, (111)-Cl                                                            & 7.10 & 3.5  & y \\ \hline

\end{tabular}
\caption{Surfaces with ultra-low and ultra-high work functions sorted by spacegroup (sg). Surfaces stemming from structurally equivalent materials (but occupied with different chemical species) are grouped together into families described by chemical formulas containing placeholder letters A and B. For each family the compound that resulted in the lowest/highest work function is listed alongside its work function value and cleavage energy. The Miller indices and the atom termination of the surfaces yielding extreme work functions are specified. Work functions are given in eV and cleavage energies in meV/$\AA{}^2$. The cleavage energy is equal to the surface energy for slabs which state y(es) in the right-most column (\textit{i.e.}, the bottom and top surfaces of these slabs are equivalent while the slab maintains stoichiometry), otherwise $2\cdot E_\mathrm{cleavage}$ is an upper bound for the surface energy.}
\label{tab:extremeWF}
\end{table*}

\subsection{Machine Learning Model for Work Function Predictions}

\begin{figure*}[tp]
\includegraphics[width=\textwidth]{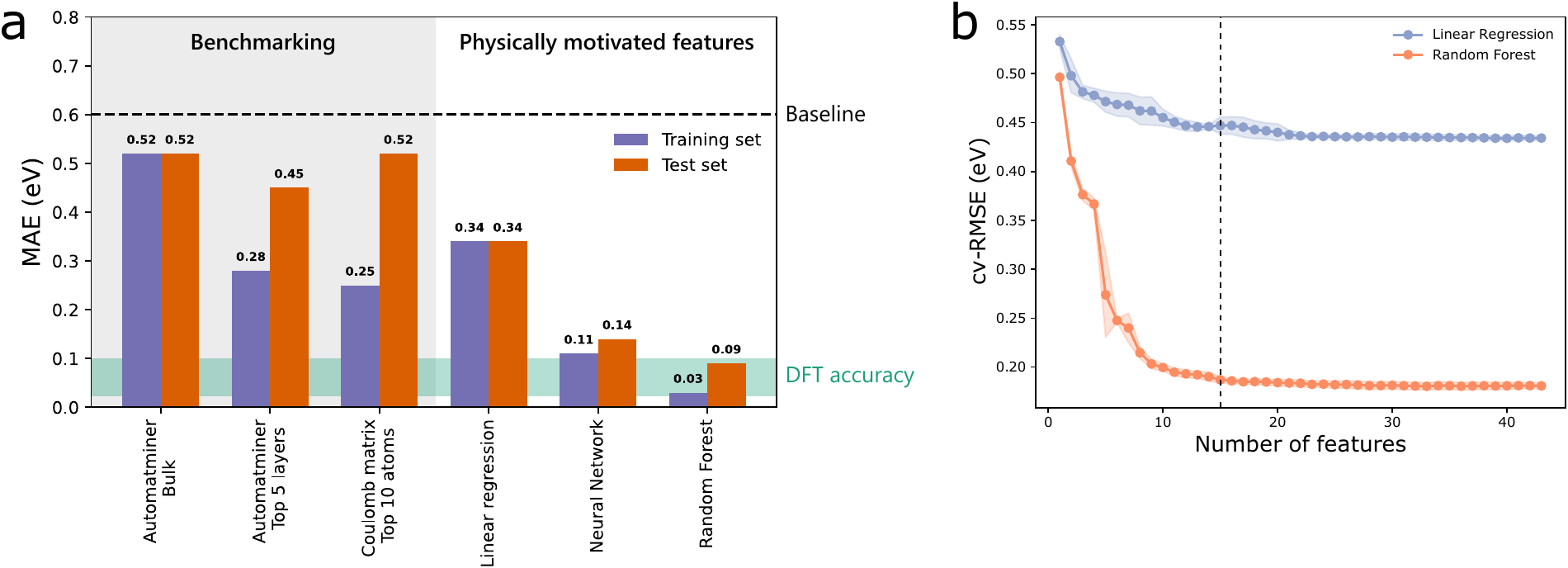}
\caption{Comparison of machine learning model performances. \textbf{a} Mean absolute errors (MAEs) of training and test sets are given for this paper's machine learning models:  Linear regression, neural network, and random forest implementing 15 physically motivated features. The benchmarking models (automatminer with bulk unit cells and with surface slab of topmost 5 atomic layers as inputs, and a Coulomb matrix of the topmost 5 surface atoms, $\ell_2$-sorted, with a random forest model) are shown in comparison. The baseline model (always guessing the average work function) and the DFT accuracy are indicated by a dashed line and green-shaded area, respectively. The upper limit of the DFT accuracy (0.1 eV) is estimated from the accuracy of the work function calculations of elemental crystals.\cite{DeWaele2016} \textbf{b} 5-fold cross-validated root mean square error (RMSE) as a function of number of input features is shown. Features were selected by recursive feature elimination for linear regression and random forest model. The top 15 most predictive features were selected for the final models.
}%
\label{fig:ML}
\end{figure*}

\begin{figure}
\includegraphics[width=0.45\textwidth]{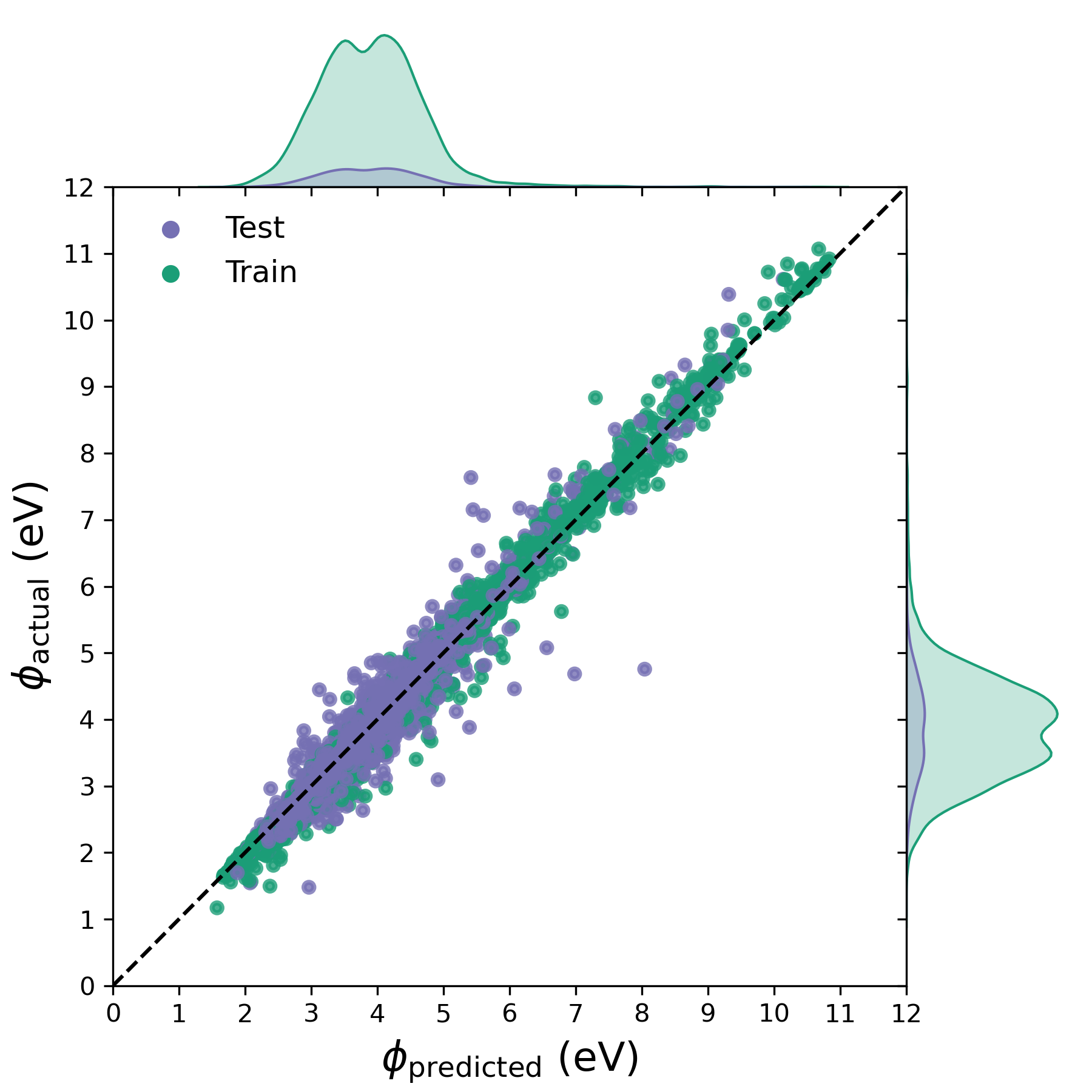}
\caption{Predicted work functions vs. DFT-calculated work functions. The kernel-density estimate distributions for both training and test sets are plotted for predicted work functions and the ground-truth at the top and right, respectively. 
}%
\label{fig:RF}
\end{figure}

The large database created by high-throughput DFT calculations forms the basis for a surrogate machine learning model that enables the prediction of the work function at a fraction of the computational cost. As a first step, we assess common models from the  materials science machine learning community as a benchmark. For that, we employ the automatminer testing suite,\cite{Dunn2020} and a conventional Coulomb matrix (trained with a random forest model).\cite{Himanen2020} For automatminer we use the ``express" setting and compare using the bulk unit cell and the topmost 5 atomic layers of the surface slabs as inputs. As a baseline model we predict the work function to be the average work function regardless of the surface. The automatminer model performs only marginally better than the baseline model when bulk structures are used as an input. When the surface slabs are used as inputs the performance increases and is comparable to the performance of the Coulomb matrix. The mean absolute errors (MAEs) are shown for the training and test sets in Figure \ref{fig:ML}a (\textit{cf.} Figure S9 for RMSEs). The baseline MAE is 0.60 eV and the DFT accuracy is indicated in the green-shaded area between 0.022 and 0.1 eV, corresponding to the convergence error (see Methods) and the error between PBE-calculated and experimental work functions,\cite{DeWaele2016} respectively.\\
It is not surprising that the model performance is poor when the bulk structure is used as an input as the database contains multiple surfaces of different work functions for any given bulk structure. While the performance of the benchmarking models improves when the surface slab is used as the input instead, the MAEs are still large and significant overfitting is observed. This is likely due to the fact, that the models cannot distinguish between top and bottom of the input slab (which in general are not symmetric) and the database consists of all unique terminations. In general, if one termination (located at the top surface) is labeled with the calculated work function, the same termination exists in another input structure at the bottom surface (whereas the calculated work function always refers to the top surface). Hence, the shortcomings of the automated benchmarking models does not come from the machine learning models used but rather the implementation of the featurization of surface slabs.\\
We developed a custom featurization of the surface slabs by considering physically motivated features of the topmost three surface layers. Atoms are regarded to belong to the same layer if their $c$-coordinate lies within a tolerance of 0.4 $\AA{}$ (see effect of tolerance value on model performance in Figure S7). The considered atomic features are electronegativity $\chi$, inverse atomic radius $1/r$, first ionization energy $E_\mathrm{ion}$, and Mendeleev number $n_\mathrm{mend}$. Given that each layer may contain more than one atom-type, we consider the minimum, maximum, and average of each of these atomic features. This gives a total of 36 elemental features for the topmost 3 layers. Additionally, we add structural features: The packing fraction for each layer (number of atoms per unit cell area) $A_\mathrm{pack}^{-1}$ and the distances between atomic layers 1 and 2, $d_{1-2}$, and between layers 1 and 3, $d_{1-3}$. Angle-based features are calculated by considering the angles between the surface normal vector and the vectors spanned between atoms in the topmost layer and their respective nearest neighbors. After all angles are calculated we consider the minimum and maximum values as two features, $\theta_\mathrm{min}$ and $\theta_\mathrm{max}$. Out of this total 43 features the most significant features are selected with (backward) recursive feature elimination (RFE) using a random forest model, as plotted in Figure \ref{fig:ML}b. The top 8 features largely account for the model performance: $\left<\chi_1\right>$, $\left<E_\mathrm{ion,1}\right>$, $\left<E_\mathrm{ion,2}\right>$, $\left<n_\mathrm{mend,1}\right>$,  $d_{1-3}$, $A_\mathrm{pack,1}^{-1}$, $\mathrm{min}\left(\chi_1\right)$, and $\mathrm{min}\left(1/r_1\right)$. For the final model we use the best 15 features, which are the 8 features mentioned above and $\theta_\mathrm{min}$, $d_{1-2}$ $\left<1/r_1\right>$, $\mathrm{min}\left(E_\mathrm{ion,2}\right)$, $\left<E_\mathrm{ion,3}\right>$, $\mathrm{max}\left(\chi_2\right)$, and $\mathrm{min}\left(n_\mathrm{mend,2}\right)$.\\
It is worth noting that the majority of features we selected were physics-motivated or based on correlations observed in literature. The work function has been shown to linearly correlate with electronegativity for elemental crystals and binary compounds,\cite{Tran2016, Yamamoto1974}, and inversely correlate with the atomic radius. Another work has proposed a phenomenological equation for the work function that depends on the atomic radius and the number of atomic sites per unit cell area at the surface.\cite{Brodie2014} We chose the ionization energy as a feature based on physical intuition that low ionization energies lead to easy electron extraction. Lastly, the Mendeleev number has been shown to be a descriptive feature for many material property predictions.\cite{Cheon2018a,Villars2004} Interestingly, the most predictive features (top 8) are features from the topmost layer (including the layer distance $d_{1-3}$) with the exception of $\left<E_\mathrm{ion,2}\right>$. This is in agreement with the the fact that clear trends are observed considering only the topmost surface (\textit{cf.} Figure \ref{fig:WF_trends}a). The final 15 features contain only two features that relate to the third atomic layer: $\left<E_\mathrm{ion,3}\right>$ and $d_{1-3}$. Also, we note that we tried adding further elemental features without a clear physics-based motivation (\textit{e.g.}, polarizability) as well as further modes (\textit{e.g.}, geometric mean, range, variance) -- however, this did not improve the model performance.\\
Using this featurization approach (with 15 features) outperforms all benchmarking models (automatminer, in comparison, uses 200 features) even when a linear regression model is chosen, as seen in Figure \ref{fig:ML}a. This performance improvement is due to the superior implementation of how we featurize our slabs rather than the machine learning model itself. When a non-linear learning model is used (neural network or random forest model) the MAEs are significantly reduced. Our best model using random forests has a test-MAE of 0.09 eV, comparable to the accuracy of work function calculations employing DFT. This test performance is about 4--5 times better than the best benchmarking model and more than 6 times better than the baseline. Figure \ref{fig:RF} shows the predicted work function for both the training and test sets in comparison to the DFT-calculated values. The kernel-density estimate distributions for both training and test sets are plotted for predicted and actual work functions showing that the predicted distribution is qualitatively faithful to the actual one. Notably, for the neural network and random forest models there is still a gap between training and test MAEs despite thorough hyperparameter tuning. The learning curves in Figure S8 indicate underfitting for the linear model and slight overfitting for the random forest model. The learning curve trend for the random forest model suggests that increasing the dataset size (by a factor of $\sim 10$) could further improve the model performance and close the gap between training and cross-validated errors.\\
The prediction of the work function using this model is roughly $10^5$ times faster than DFT while having a MAE comparable to the accuracy of DFT. The database and model are available open-source (see \textit{data availability} section) enabling other researchers to use this model for work function predictions and help experimentalists in their materials/synthesis choice.

\section{Conclusions}
In summary, we demonstrate a workflow to create a work function and cleavage energy database from high-throughput DFT calculations that enables us to gain insight into chemical trends of the work function and surface stability. The database reveals 34 ultra-low and 56 ultra-high work function surfaces many of which have low surface energies. The low surface energy candidates are expected to be the stable surface termination and orientation of the corresponding stable bulk material and are hence promising candidates for experimental verification. To our knowledge, we identify the lowest and highest computationally predicted work functions in literature (1.25 and 9.06 eV, respectively) for steady-state metallic surfaces without requiring any monolayer coatings. Further, we establish a surrogate machine learning model for rapid work function predictions. Our model has a MAE of 0.09 eV, comparable to the accuracy of DFT while being $\sim10^5$ times faster. Using this approach facilitates the probing of a vast chemical space for novel material surfaces with exceptionally low or high work functions.

\section{Author Information}
\subsection{Corresponding author} *E-mail:\\p.schindler@northeastern.edu

\subsection{Data Availability}
The database is available in a json format at \href{https://doi.org/10.5281/zenodo.10381506}{10.5281/zenodo.10381506} and the machine learning model is available at \href{https://github.com/d2r2group/WF-RF-Model}{github.com/d2r2group/WF-RF-Model} and v2.1.1 of the code is also available at \href{https://doi.org/10.5281/zenodo.10449568}{10.5281/zenodo.10449568}.

\subsection{Author Contributions}
\textbf{Peter Schindler}: Writing – original draft (lead); Review and editing (equal); Software (lead); Methodology (lead); Visualization (lead); Investigation (lead); Data curation (lead); Conceptualization (supporting);  Supervision (equal); Resources (lead). \textbf{Evan R. Antoniuk:} Review and editing (equal); Methodology (supporting); Conceptualization (supporting). \textbf{Gowoon Cheon:} Review and editing (equal); Conceptualization (supporting). \textbf{Yanbing Zhu:} Review and editing (equal); Conceptualization (supporting). \textbf{Evan J. Read:} Conceptualization (lead); Review and editing (equal); Supervision (equal)

\section{Acknowledgement}
P.S. gratefully acknowledges the start-up funds from Northeastern University, Department of Mechanical and Industrial Engineering. This work was completed in part using the Discovery cluster, supported by Northeastern University's Research Computing team. Author and Professor Evan J. Reed passed away March 2022. He contributed great guidance toward the project. We are incredibly grateful for his contributions.

\bibliography{library}
\clearpage
\includepdf[pages={1}]{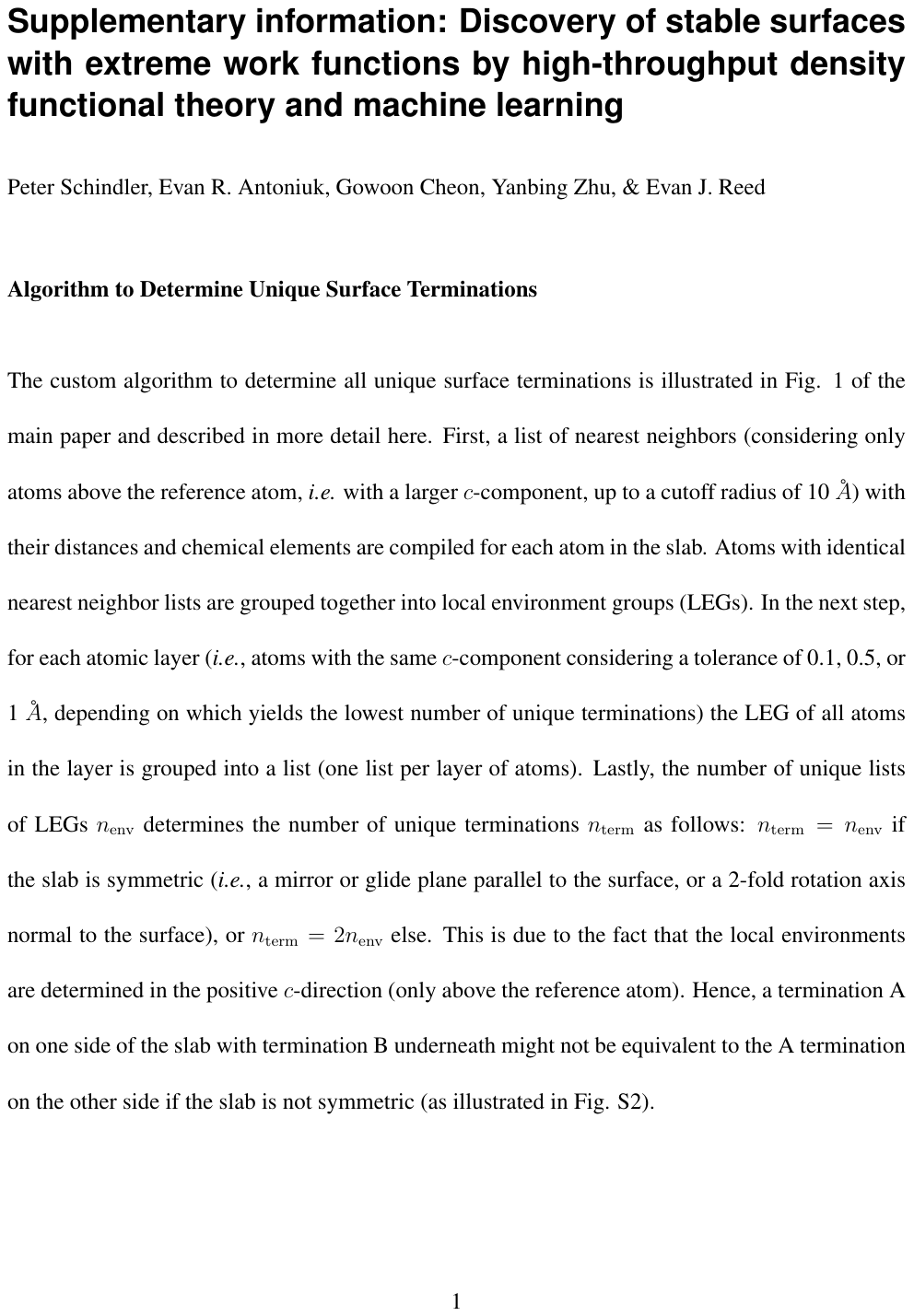}
\clearpage
\includepdf[pages={2}]{Suppl.pdf}
\clearpage
\includepdf[pages={3}]{Suppl.pdf}
\clearpage
\includepdf[pages={4}]{Suppl.pdf}
\clearpage
\includepdf[pages={5}]{Suppl.pdf}
\clearpage
\includepdf[pages={6}]{Suppl.pdf}
\clearpage
\includepdf[pages={7}]{Suppl.pdf}
\clearpage
\includepdf[pages={8}]{Suppl.pdf}
\clearpage
\includepdf[pages={9}]{Suppl.pdf}
\clearpage
\includepdf[pages={10}]{Suppl.pdf}
\clearpage
\includepdf[pages={11}]{Suppl.pdf}
\clearpage
\includepdf[pages={12}]{Suppl.pdf}
\clearpage
\includepdf[pages={13}]{Suppl.pdf}
\clearpage
\includepdf[pages={14}]{Suppl.pdf}
\clearpage
\includepdf[pages={15}]{Suppl.pdf}
\clearpage
\includepdf[pages={16}]{Suppl.pdf}
\clearpage
\includepdf[pages={17}]{Suppl.pdf}
\clearpage
\includepdf[pages={18}]{Suppl.pdf}
\end{document}


\maketitle
\section*{Algorithm to Determine Unique Surface Terminations}
The custom algorithm to determine all unique surface terminations is illustrated in Fig. 1 of the main paper and described in more detail here. First, a list of nearest neighbors (considering only atoms above the reference atom, \textit{i.e.} with a larger $c$-component, up to a cutoff radius of 10 $\AA{}$) with their distances and chemical elements are compiled for each atom in the slab. Atoms with identical nearest neighbor lists are grouped together into local environment groups (LEGs). In the next step, for each atomic layer (\textit{i.e.}, atoms with the same $c$-component considering a tolerance of 0.1, 0.5, or 1 $\AA{}$, depending on which yields the lowest number of unique terminations) the LEG of all atoms in the layer is grouped into a list (one list per layer of atoms). Lastly, the number of unique lists of LEGs $n_\mathrm{env}$ determines the number of unique terminations $n_\mathrm{term}$ as follows:  $n_\mathrm{term}= n_\mathrm{env}$ if the slab is symmetric (\textit{i.e.}, a mirror or glide plane parallel to the surface, or a 2-fold rotation axis normal to the surface), or $n_\mathrm{term}=2 n_\mathrm{env}$ else. This is due to the fact that the local environments are determined in the positive $c$-direction (only above the reference atom). Hence, a termination A on one side of the slab with termination B underneath might not be equivalent to the A termination on the other side if the slab is not symmetric (as illustrated in Fig. S\ref{fig:2d}).

\section*{Previous Version of Work Function Database}
In a previous version of the database we included materials with a non-zero band gap during screening and trained machine learning models on it. However, eventually this database was not utilized in the final paper as band gaps for PBE-level DFT calculations are well-known to be unreliable and hence the work function (positioned at the center of the gap) may also suffer from the same limitation. We do however note that this error is potentially small for materials with a small band gap (less than about half an eV) with a potential partial error cancellation w.r.t.\ the vacuum energy level. A demonstration of that is the fact that we were able to identify low work function surfaces in non-metallic materials such alkali oxides (\textit{e.g.}, Cs$_2$O$_2$), alkaline vanadates (\textit{e.g.}, BaVO$_3$), and alkali-based anti-fluorite structures (\textit{e.g.}, Rb$_2$O, Rb$_2$S) that were already known in literature (\textit{cf.} introduction in the main paper). The lowest work function after ionic relaxation was calculated for the (100)-Rb surface of PdRb$_2$C$_2$ at 0.93 eV.\\
Most high work function candidates in the previous version of the database exhibit a larger bandgap, as expected for flourides, carbides, nitrides and oxides. However, there are a few materials with a zero PBE bangdap in spacegroups 66 (AgO), 139 (CsO$_2$, RbO$_2$, KO$_2$, CaN$_2$), 221 (ReO$_3$), and 225 (PbO$_2$, BiO$_2$). The highest work function after ionic relaxation was calculated for the (110)-F surface of AlF$_3$ at 11.04 eV, likely not in agreement with experimental work functions due to the large band gap error in PBE.\\
The previous version of the database and the machine learning models trained on that database are available at \textbf{https://github.com/peterschindler/WorkFunctionDatabase} .

\bibliography{../library.bib}
\clearpage

\begin{figure}
\includegraphics[width=\textwidth]{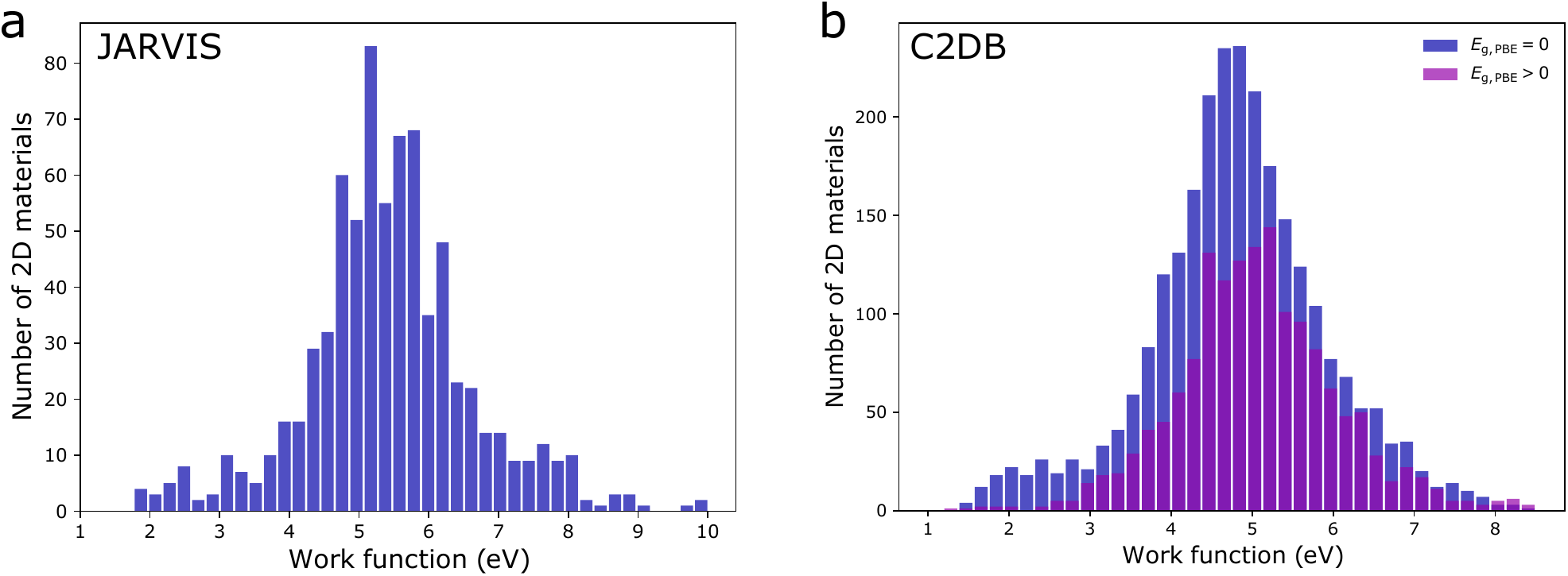} 
\caption{\textbf{Work function distributions of 2-dimensional material databases.} Work functions distributions are plotted for \textbf{a} 2D materials of JARVIS-DFT database and \textbf{b} C2DB (metallic and semiconducting 2D materials are indicated in purple and magenta, respectively).}%
\label{fig:2d}
\end{figure}
\begin{table}
\begin{tabular}{c|cccccc}
\hline
\textbf{Database} &\textbf{Avg/St.Dev.}& \textbf{Min/Max} & \parbox{4cm}{\centering\textbf{Material IDs}\\[-0.5\baselineskip] \textbf{with} \boldmath$\phi<2$ \textbf{eV}} & \parbox{4cm}{\centering\textbf{Material IDs}\\[-0.5\baselineskip] \textbf{with} \boldmath$\phi>8$ \textbf{eV}}\\ 
\hline
JARVIS & $5.43\pm 1.22$ & $1.8/10.0$ & 
\parbox{4cm}{\tiny\centering JVASP-8942, JVASP-9002,\\[-0.5\baselineskip] JVASP-9059, JVASP-9065,\\[-0.5\baselineskip] JVASP-19991} & 
\parbox{4cm}{\vspace{.3\baselineskip}\tiny\centering JVASP-783, JVASP-765,\\[-0.5\baselineskip] JVASP-31368, JVASP-31379,\\[-0.5\baselineskip] JVASP-14441, JVASP-60555,\\[-0.5\baselineskip] JVASP-14456, JVASP-14458,\\[-0.5\baselineskip] JVASP-75058, JVASP-75066,\\[-0.5\baselineskip] JVASP-28011, JVASP-75269,\\[-0.5\baselineskip] JVASP-28212, JVASP-60599,\\[-0.5\baselineskip] JVASP-14460, JVASP-60359,\\[-0.5\baselineskip] JVASP-6244, JVASP-6277,\\[-0.5\baselineskip] JVASP-27755, JVASP-28153,\\[-0.5\baselineskip] JVASP-60236, JVASP-75154,\\[-0.5\baselineskip] JVASP-60593\vspace{.5\baselineskip}}
\\
\hline
C2DB & $4.91\pm 1.08$ & $1.4/9.5$ & 
\parbox{4cm}{\vspace{.3\baselineskip}\tiny\centering B2H2O2Zr3-b82e8dae99dd,\\[-0.5\baselineskip] C2H2O2Nb3-137a187a149c,\\[-0.5\baselineskip] CH2O2V2-c08bda91646b,\\[-0.5\baselineskip] C2H2O2Sc3-c6f3a447a0d1,\\[-0.5\baselineskip] H2N2O2Mn3-46340d3ecfb9,\\[-0.5\baselineskip] H2O2N3Mn4-0575209d5cde,\\[-0.5\baselineskip] MnH2O2-3bd136c8a265,\\[-0.5\baselineskip] OsH2O2-192c92216c12,\\[-0.5\baselineskip] NCr2H2O2-ff61539f0f78,\\[-0.5\baselineskip] H2O2C3Cr4-6b206ce0e0a4,\\[-0.5\baselineskip] H2O2N3Mo4-f9b42b34c5ab,\\[-0.5\baselineskip] CoH2O2-ebe1b342d92f,\\[-0.5\baselineskip] CH2O2Y2-310f97f320f8,\\[-0.5\baselineskip] C2H2O2Ti3-b2d22251cb52,\\[-0.5\baselineskip] NH2Mn2O2-af38ffeefb90,\\[-0.5\baselineskip] C2H2O2V3-9c6994bf71a2,\\[-0.5\baselineskip] H2N2O2Sc3-3c2898c13b0f,\\[-0.5\baselineskip] H2O2C3Mn4-200df1168731,\\[-0.5\baselineskip] H2O2C3Mo4-2e0b9178f9f2,\\[-0.5\baselineskip] FeH2O2-35be111741a4,\\[-0.5\baselineskip] RhH2O2-44bcea466c91,\\[-0.5\baselineskip] C2H2O2Y3-cbf305eb62b1,\\[-0.5\baselineskip] H2O2C3Nb4-760d7905c535,\\[-0.5\baselineskip] GaH2O2-b42b47e92444,\\[-0.5\baselineskip] RuH2O2-cd1e59aa5b1c,\\[-0.5\baselineskip] PdH2Li2-823b99472212,\\[-0.5\baselineskip] Te2Zn2N4H8-f34362b9488b,\\[-0.5\baselineskip] H2O2C3Sc4-33474bb0af08,\\[-0.5\baselineskip] H2O2N3V4-f09ac233cc92,\\[-0.5\baselineskip] CoCa2O3-40b5f2e2e758,\\[-0.5\baselineskip] C2H2S2Ti3-4cada7c36f9f,\\[-0.5\baselineskip] H2N2O2Y3-9030574e723a,\\[-0.5\baselineskip] H2O2C3Ti4-ec020ed8f687,\\[-0.5\baselineskip] NiH2O2-4f81e6b4f5aa,\\[-0.5\baselineskip] CH2Mn2O2-7938aedd38e3,\\[-0.5\baselineskip] C2H2S2Zr3-2abe5cf3179c,\\[-0.5\baselineskip] H2O2C3V4-8f5f4356ee52,\\[-0.5\baselineskip] C2H2O2Cr3-6996d2b04358,\\[-0.5\baselineskip] H2O2C3W4-4e95b5b75c69,\\[-0.5\baselineskip] CN-2bae1dfe5219,\\[-0.5\baselineskip] InH2O2-c58c5def89b7,\\[-0.5\baselineskip] H2O2C3Y4-82f540c2be44,\\[-0.5\baselineskip] CH2O2Sc2-f72d84df3799,\\[-0.5\baselineskip] B2H2O2Ti3-7ef79d62d408,\\[-0.5\baselineskip] C2H2O2Mn3-aae848e42008,\\[-0.5\baselineskip] B2H2O2V3-b78d257a108a,\\[-0.5\baselineskip] C2H2O2Mo3-0cde9e26013d,\\[-0.5\baselineskip] H2O2N3Cr4-900cb08d8065,\\[-0.5\baselineskip] CH2O2Ti2-b2d548e0d08a\vspace{.5\baselineskip}} & 
\parbox{4cm}{\tiny\centering Cu2F2-c4860f83a9b8,\\[-0.5\baselineskip] GaO2-0a65a6b05ce6,\\[-0.5\baselineskip] BaTa2O7-877ae50c1f18,\\[-0.5\baselineskip] SnO2-d5f47e5d4cf7,\\[-0.5\baselineskip] PbF4-0ace524dc18a,\\[-0.5\baselineskip] SnF4-a61b85728555,\\[-0.5\baselineskip] F2Os2-63abc51956be,\\[-0.5\baselineskip] PbO2-0963dec75c0c,\\[-0.5\baselineskip] Al2O4-b003ee55f237,\\[-0.5\baselineskip] BaSb2F12-a8ad6f8ad8ee,\\[-0.5\baselineskip] F2Ir2-6ab59c2ae465,\\[-0.5\baselineskip] O4V4F12-05e5d7a8d56c,\\[-0.5\baselineskip] PdF2-ee10b74aa014,\\[-0.5\baselineskip] AgSnF6-edbdca5bd78a,\\[-0.5\baselineskip] AlFeF5-2c9ff5807948,\\[-0.5\baselineskip] SrTa2O7-a4995476d4ae,\\[-0.5\baselineskip] C2Ca2F2O6-b46e73deeb64,\\[-0.5\baselineskip] Ti4O8-92c270a35817,\\[-0.5\baselineskip] CaAu2F12-73f67052a2e7,\\[-0.5\baselineskip] F2N2O2Zr3-57adb92deade,\\[-0.5\baselineskip] RhF2-3574951e1edf
} \\ 
\hline
\end{tabular}
\caption{\linespread{1}\selectfont{}Detailed work function distribution metrics for 2D materials in JARVIS-DFT database and C2DB. All values in eV. Unique material IDs are listed for materials with work functions below 2 and above 8 eV.} 
\label{tab:distr}
\end{table}
\begin{figure}
\includegraphics[width=0.5\textwidth]{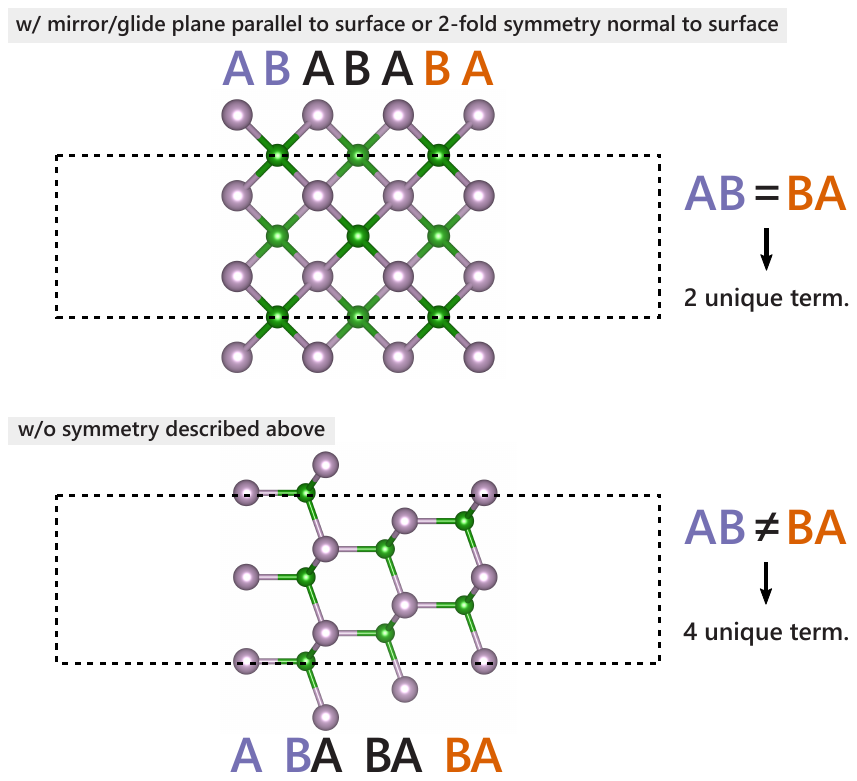} 
\caption{\textbf{Method section illustrations.} Illustration of the determination of the effective number of unique terminations based on symmetry of the slab. The two slabs both have two unique local environment groups (A and B) and correspond to 2 and 4 unique terminations for a symmetric slab in $c$-direction and a non-symmetric slab, respectively.}%
\label{fig:2d}
\end{figure}
\begin{figure}
\includegraphics[width=\textwidth]{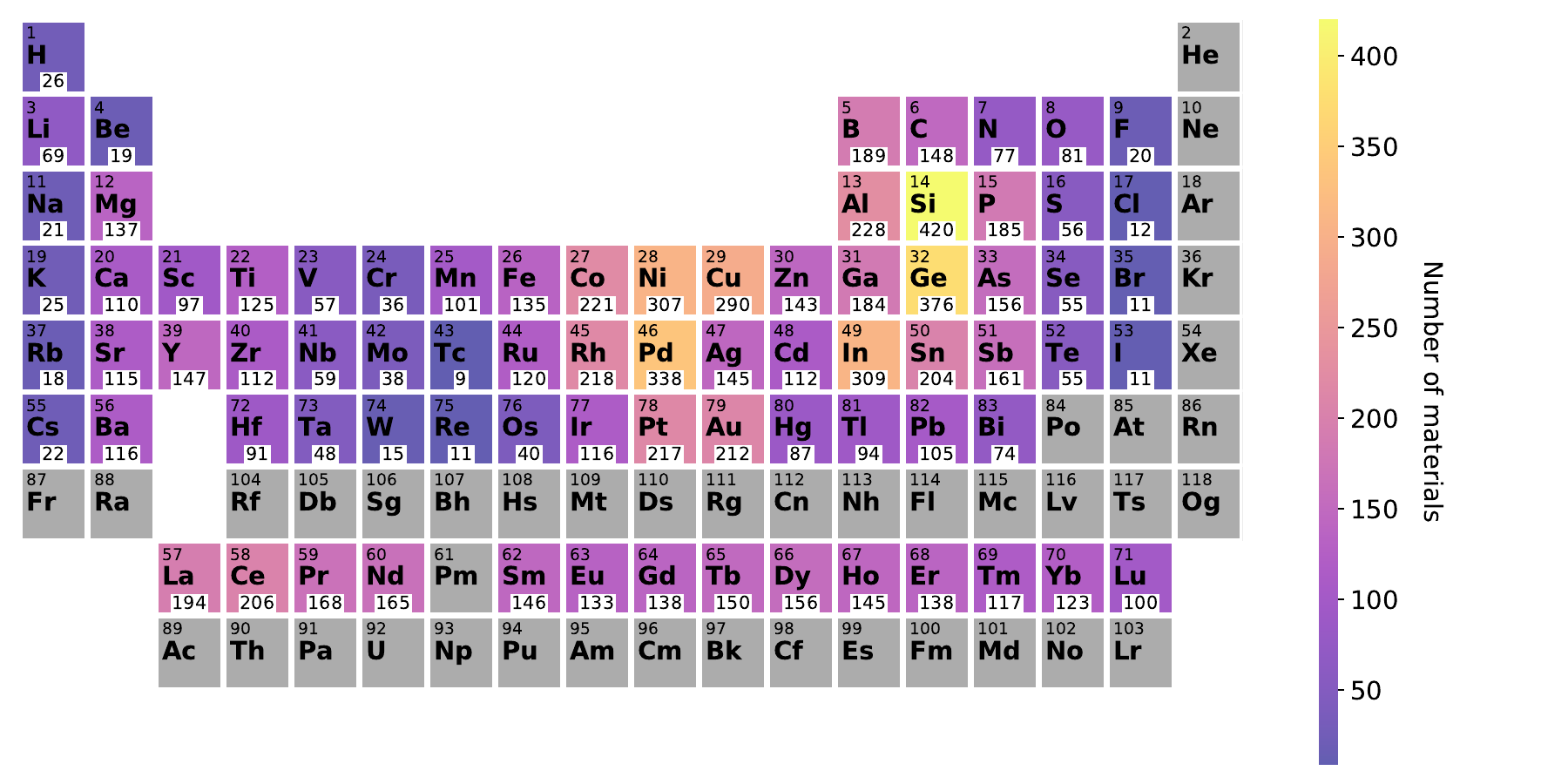} 
\caption{\textbf{Heat-map of elements present in the bulk compounds used to create the work function database.} Si, Ge, and Pd are the most common elements, whereas Tc, Re, Br and I are the least common.}%
\label{fig:periodictable}
\end{figure}
\begin{figure}
\includegraphics[width=\textwidth]{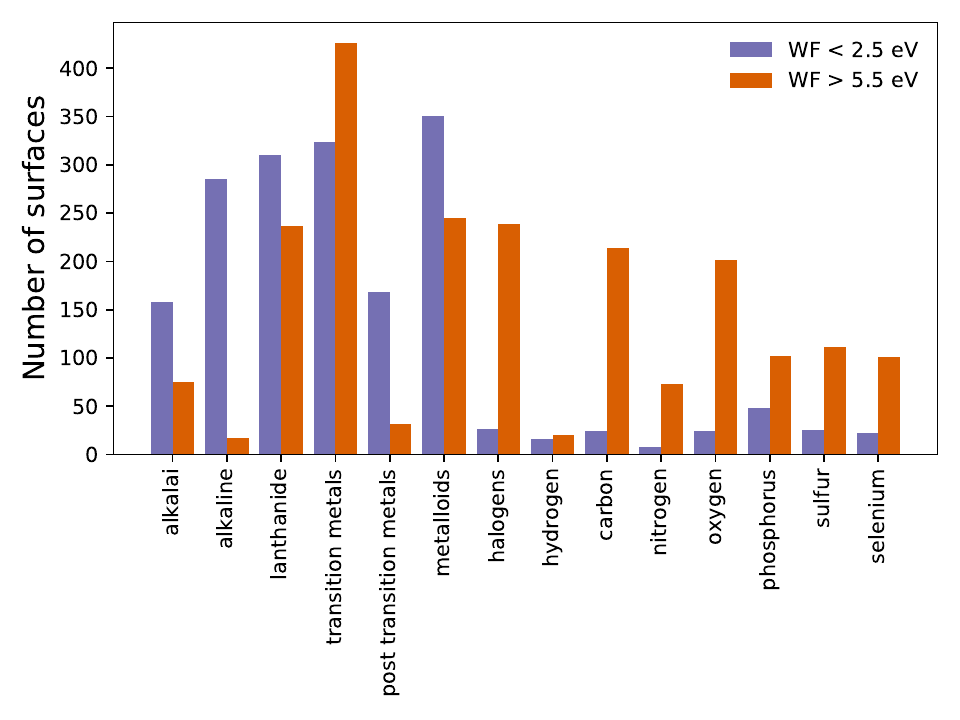} 
\caption{\textbf{Total number of surfaces with low and high work functions plotted as a function of chemical species present at the topmost layer.} Analogous to Figure 3a of the main text.}%
\label{fig:1layertotal}
\end{figure}
\begin{figure}
\includegraphics[width=\textwidth]{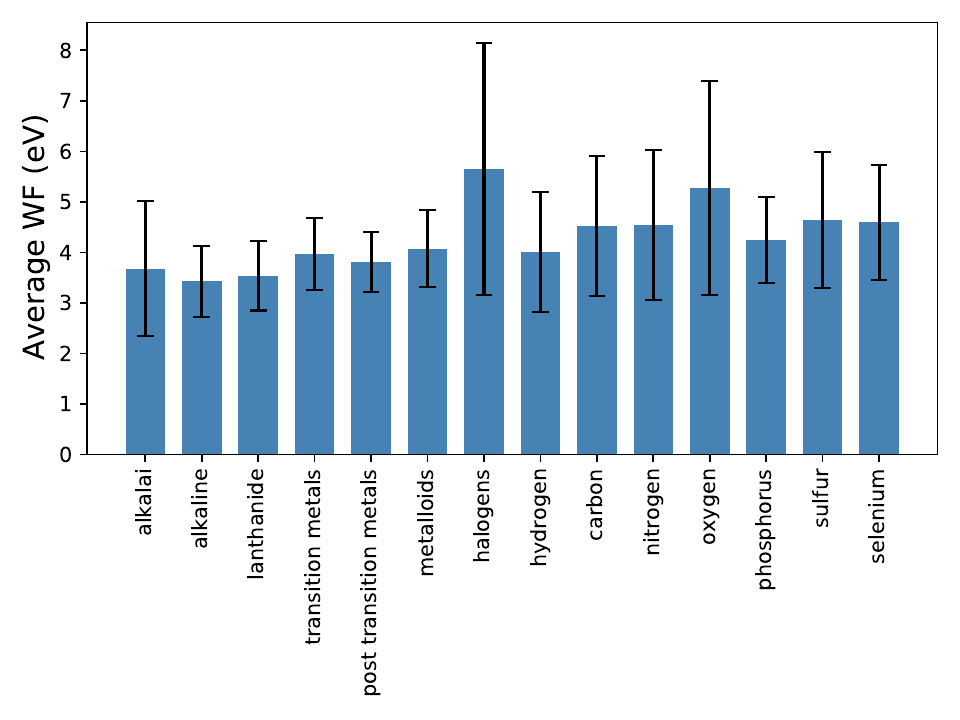} 
\caption{\textbf{Work function averages plotted as a function of chemical species present at the topmost layer.} Error bars indicate the standard deviation.}%
\label{fig:1layeravg}
\end{figure}
\begin{figure}
\includegraphics[width=0.8\textwidth]{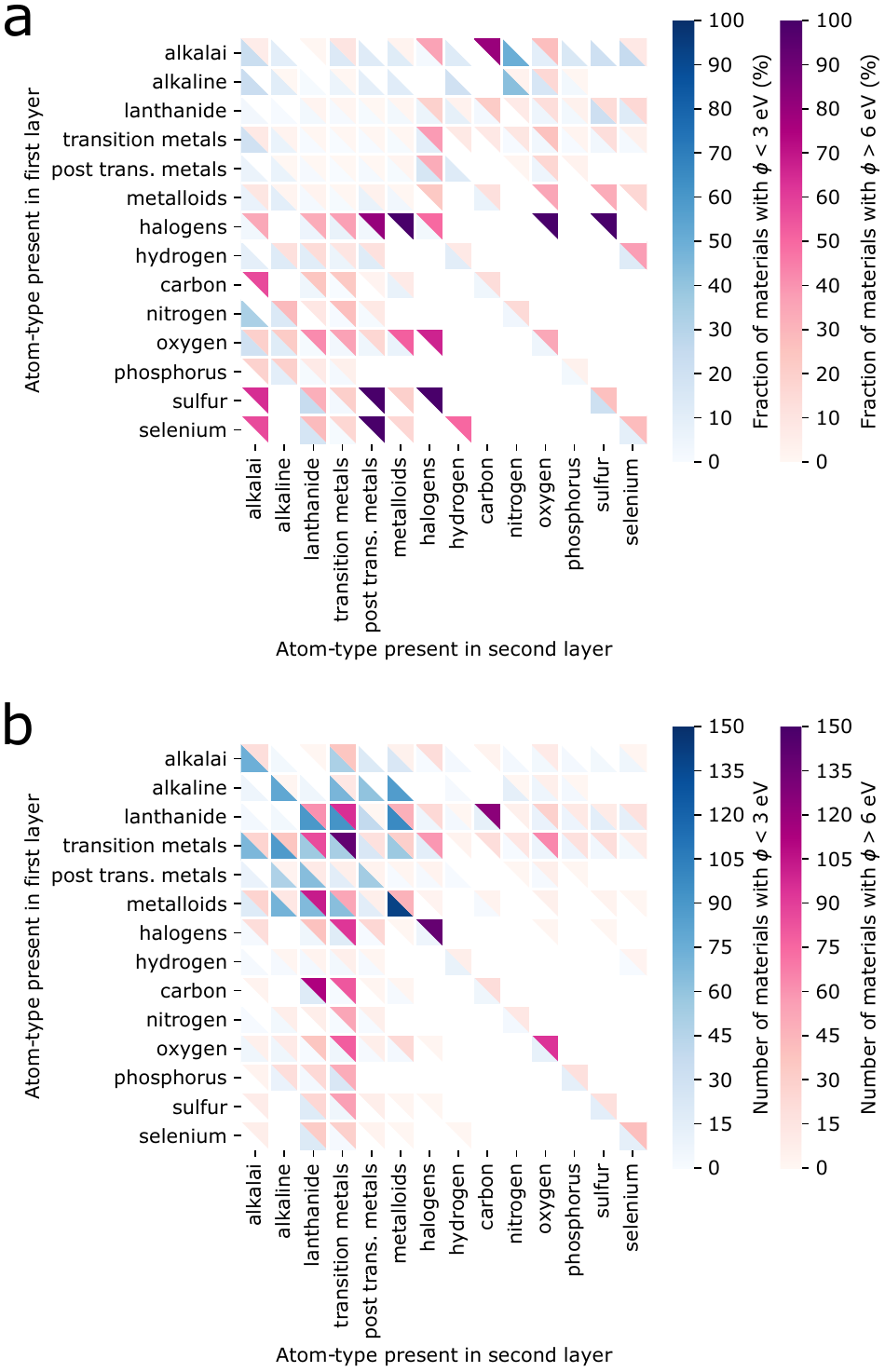} 
\caption{\textbf{Heat-maps of surfaces with low and high work functions based on chemical species present at topmost and second atomic layers.} Heat-maps plotted \textbf{a} as fractions, and \textbf{b} as total numbers.}%
\label{fig:heatmaps}
\end{figure}
\begin{figure}
\includegraphics[width=\textwidth]{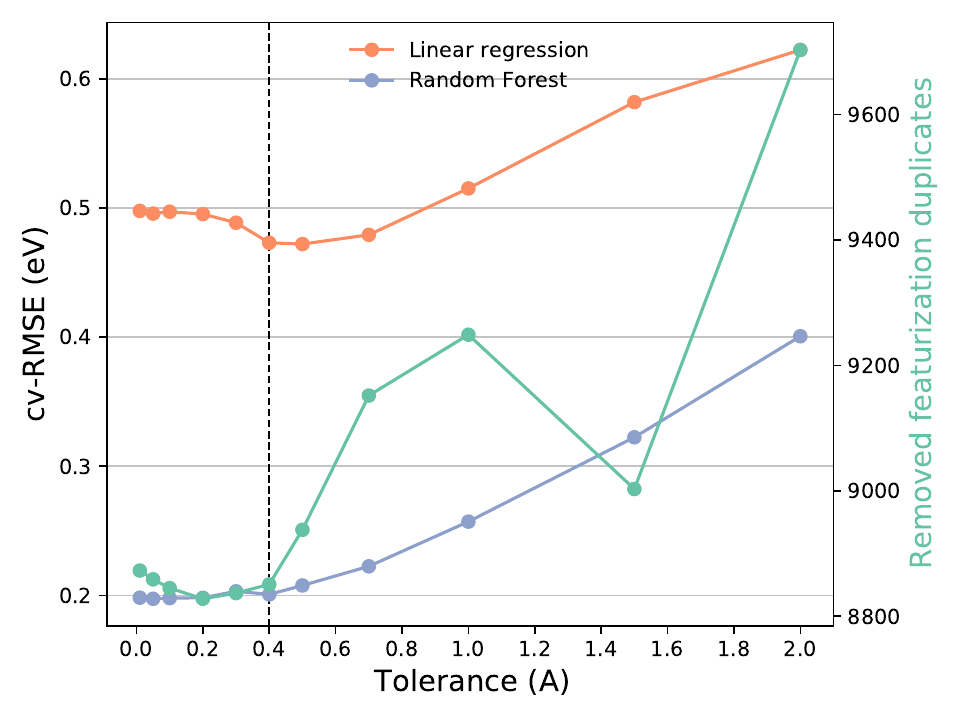} 
\caption{\textbf{RMSE plotted as a function of tolerance used for grouping atoms into layers.} 5-fold cross-validation is used for the RMSE and is shown for linear regression and random forest models. The number of feature duplicates across the database at different tolerance values is plotted in green and these surfaces are removed from the dataset before training. The final tolerance value used for models in the main paper is 0.4 $\AA{}$.}%
\label{fig:tol}
\end{figure}
\begin{figure}
\includegraphics[width=\textwidth]{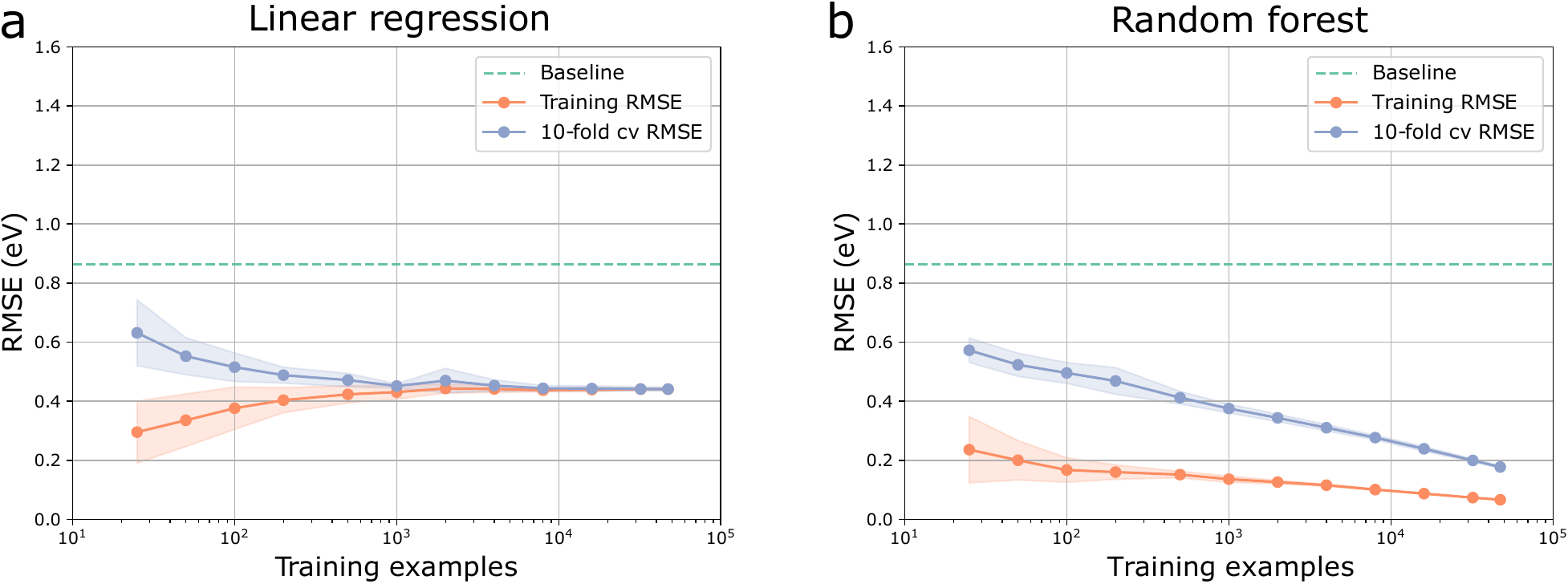} 
\caption{\textbf{Learning curves for linear regression and random forest models.} Baseline, training and 10-fold cross-validated RMSE are plotted as a function of training set size for \textbf{a} the linear regression model and \textbf{b} the random forest model. Both models used the top 15 physically-motivated features as described in the main text.}%
\label{fig:learning_curves}
\end{figure}
\begin{figure}
\includegraphics[width=\textwidth]{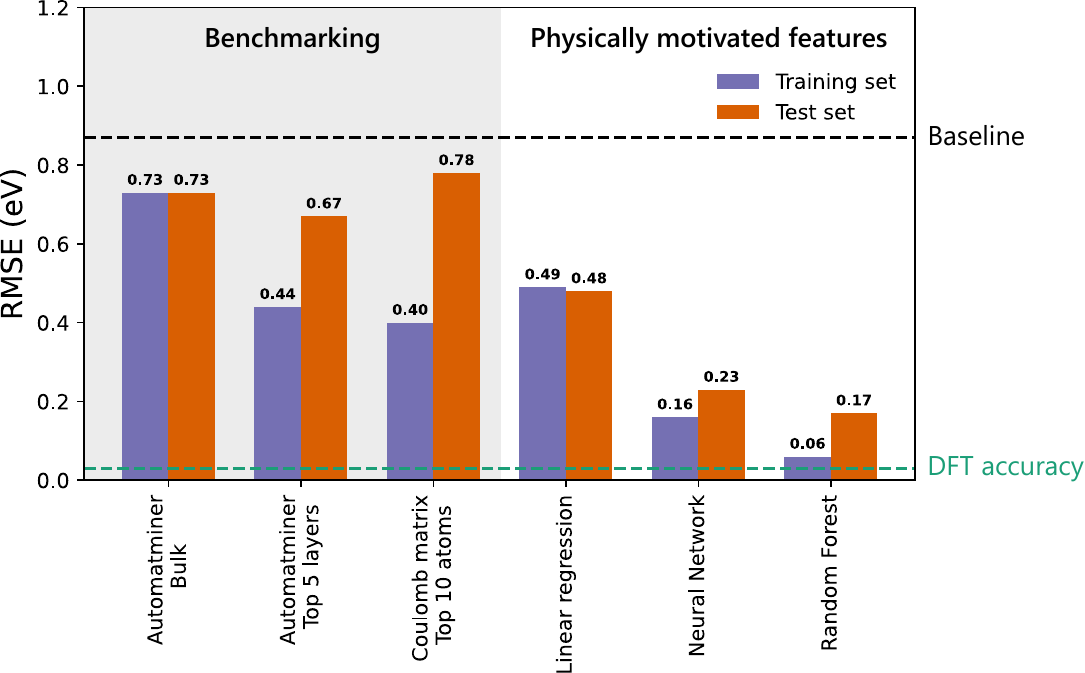} 
\caption{\textbf{Comparison of machine learning model RMSEs.} RMSEs of training and test sets are given for the machine learning models desrcibed in the main text: Linear regression, neural network, and random forest implementing 15 physically motivated features. The benchmarking models  are shown in comparison. The baseline model error of 0.87 eV (always guessing the average work function) and the DFT accuracy of 0.03 eV (lower bound) are indicated by a black and green dashed line, respectively.}%
\label{fig:results_RMSE}
\end{figure}
\begin{figure}
\includegraphics[width=\textwidth]{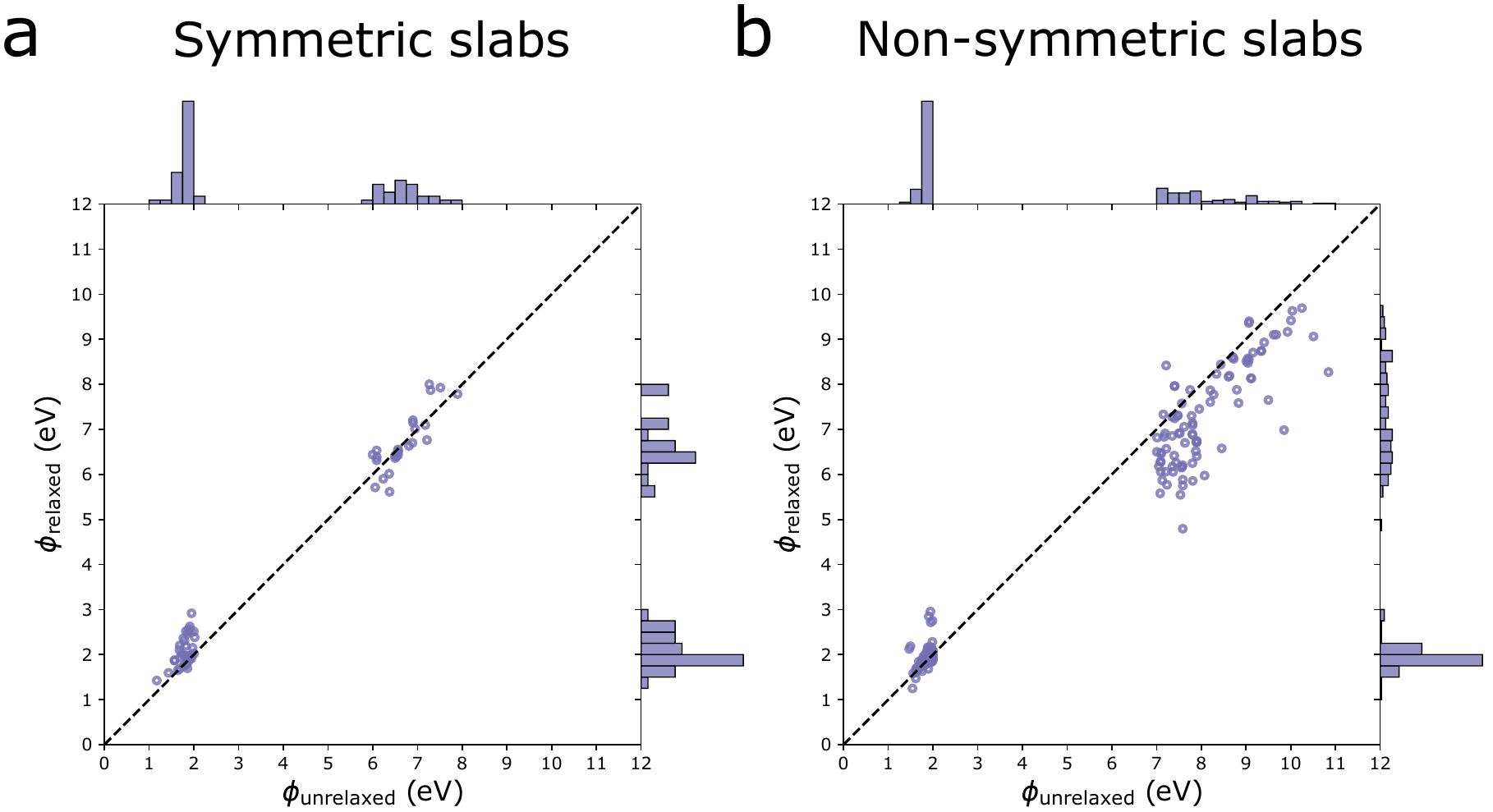} 
\caption{\textbf{Changes in work function during ionic relaxation.} Work functions of unrelaxed \textbf{a} symmetric and \textbf{b} non-symmetric slabs vs.\ the work functions of relaxed slabs are plotted for the tail ends of the unrelaxed work function distribution. Histograms are displayed to illustrate the work function distribution before and after relaxation.}%
\label{fig:tol}
\end{figure}
\begin{figure}
\includegraphics[width=0.9\textwidth]{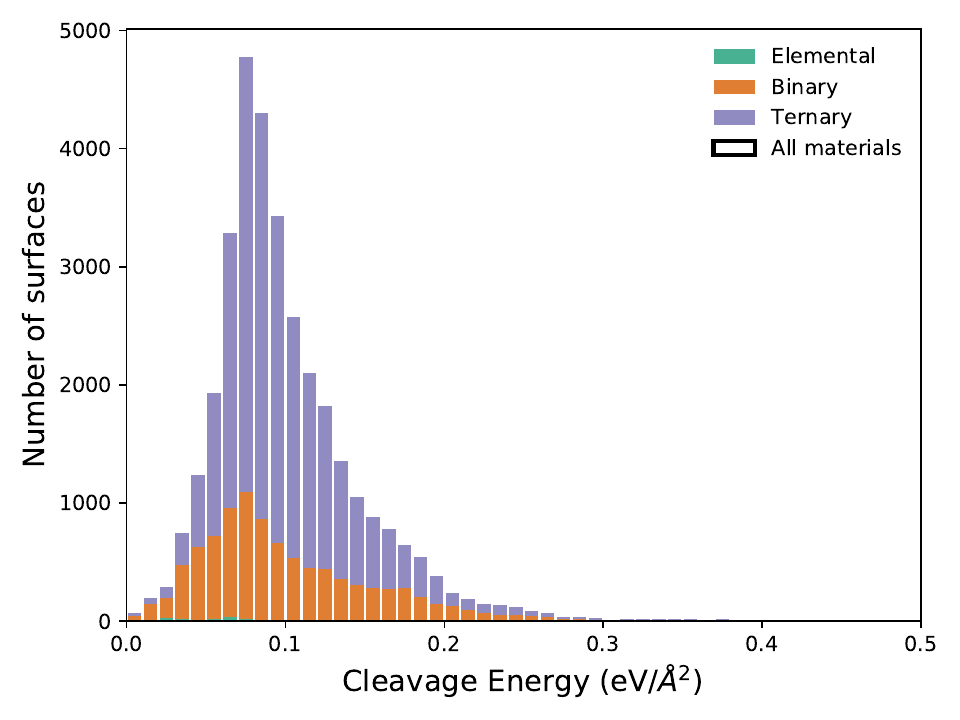} 
\caption{\textbf{Distribution of the cleavage energy plotted as a stacked histogram.} Outline corresponds to overall distribution under which the stacked, colored bars signify the number of surfaces based on elemental, binary, and ternary compounds. The average of the distribution is 101.7 meV$/\AA{}^2$ with a standard deviation of 45.7 meV$/\AA{}^2$}%
\label{fig:tol}
\end{figure}
\begin{figure}
\includegraphics[width=0.85\textwidth]{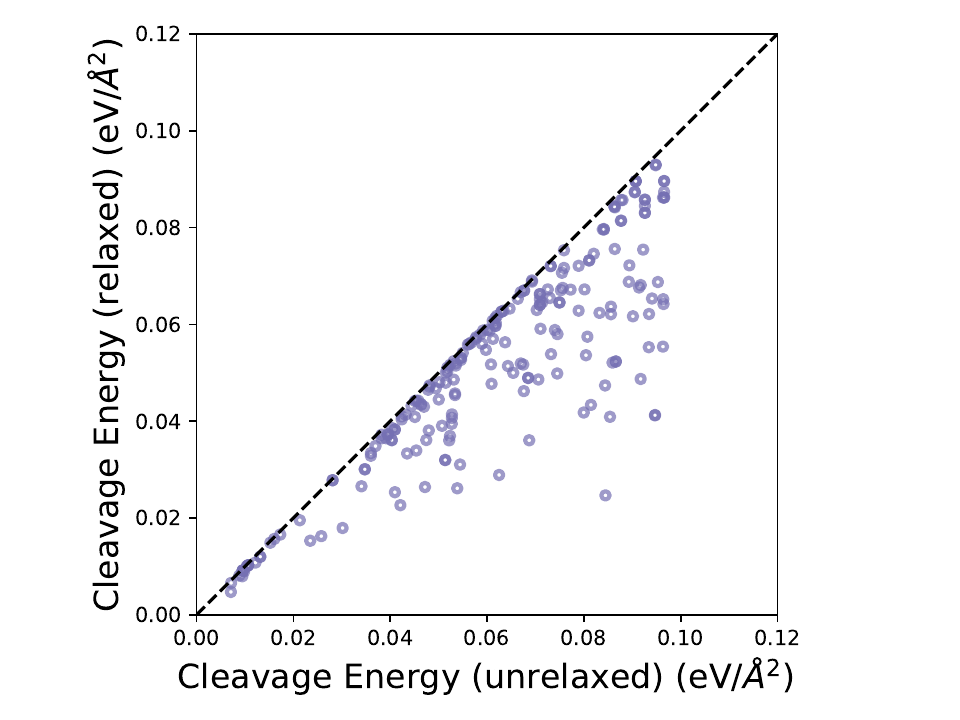} 
\caption{\textbf{Changes in cleavage energy during ionic relaxation.} Cleavage energies of unrelaxed slabs vs.\ relaxed slabs are plotted.}%
\label{fig:tol}
\end{figure}




\begin{table}
\tiny
\begin{tabular}{llllllll}
\textbf{sg} & \multicolumn{2}{l}{\textbf{Material Family}}                                       & \parbox{1cm}{\textbf{Comp. w/}\\[-0.5\baselineskip] \textbf{lowest} \boldmath$\phi$} & \parbox{2cm}{\textbf{Miller planes -}\\[-0.5\baselineskip] \textbf{Termination}  } & \boldmath$\phi$ & \textbf{mpid (mp-...)}                                                                                                                                                               & \boldmath$E_\mathrm{g}$   \\ \hline
8          &        &                                                                         & KNO$_2$                       & (100)-N                       & 2.30       & 34857                                                                                                                                                           & 2.5       \\ \cline{2-8} 
           &        &                                                                         & KCN                        & (101)-C+K                     & 1.21      & 20134                                                                                                                                                           & 5.1      \\ \hline
12         &        &                                                                         & BaNi$_2$As$_2$                   & (110)-Ba                      & 1.91      & 1070400                                                                                                                                                         & 0         \\ \cline{2-8} 
           & \textcolor{purple}{A}$_2$CN$_2$  & \textcolor{purple}{A}=Na,K                                                                  & K$_2$CN$_2$                      & (001)-K, ({\sansmath $1\bar{1}1$})-Na            & 1.29      & 10408, 541989                                                                                                                                                & 3.1 \\ \hline
38         &        &                                                                         & BaTiO$_3$                     & (001)-O-Ba                    & 1.24      & 5777                                                                                                                                                            & 2.4      \\ \hline
44         &        &                                                                         & NaNO$_2$                      & (011)-Na                      & 1.22      & 2964                                                                                                                                                            & 2.5      \\ \hline
63         & \textcolor{purple}{A}\textcolor{blue}{B}     & \textcolor{purple}{A}=Sr,Ba, \textcolor{blue}{B}=Si,Ge,Sn,Pb                                                  & BaSi                       & (110)-\textcolor{purple}{A}, (010)-\textcolor{purple}{A}              & 1.60       & \parbox{3cm}{20136,872,1730,\\[-0.5\baselineskip]2499,2661,2147}                                                                                                                    & 0         \\ \cline{2-8} 
           &        &                                                                         & KI                         & (101)-K                       & 1.74      & 1078836                                                                                                                                                         & 3.8      \\ \hline
71         & \textcolor{purple}{A}S     & \textcolor{purple}{A}=Rb,Cs                                                                 & CsS                        & (101)-\textcolor{purple}{A}                       & 1.89      & 29266,558071                                                                                                                                                 & 1.7      \\ \cline{2-8} 
           & \textcolor{purple}{A}$_2$O$_2$   & \textcolor{purple}{A}=Rb,Cs                                                                 & Cs$_2$O$_2$                      & (101)-\textcolor{purple}{A}                       & 1.12      & 7895,7896                                                                                                                                                    & 1.7-1.8 \\ \cline{2-8} 
           & \textcolor{purple}{A}\textcolor{blue}{B}$_2$\textcolor{teal}{D}$_2$  & \textcolor{purple}{A}=Pd,Pt, \textcolor{blue}{B}=K,Rb, \textcolor{teal}{D}=O,S,Se,Te                                            & PtRb$_2$S$_2$                    & (110)-\textcolor{blue}{B}, (110)-\textcolor{blue}{B}+\textcolor{teal}{D}            & 1.62      & \parbox{3cm}{\vspace{.1\baselineskip}8622,7929,8621,1068813,\\[-0.5\baselineskip]7928,540584,8623,1070356,\\[-0.5\baselineskip]1070498,1069706,1068941}                                                           & 0.7-1.4 \\ \cline{2-8} 
           & \textcolor{purple}{A}$_2$\textcolor{blue}{B}$_3$   & \textcolor{purple}{A}=Rb,Ba, \textcolor{blue}{B}=Au,Bi                                                        & Rb$_2$Au$_3$                     & (101)-\textcolor{purple}{A}                       & 1.80       & 569529,11814                                                                                                                                                 & 0         \\ \cline{2-8} 
           & Pt\textcolor{purple}{A}$_2$\textcolor{blue}{B}$_2$ & \textcolor{purple}{A}=Li,Na, \textcolor{blue}{B}=H,O                                                          & PtLi$_2$H$_2$                    & (001)-H+Li, (011)-Na          & 1.79      & 644136,22313                                                                                                                                                 & 0-1.5    \\ \hline
107        & \textcolor{purple}{A}\textcolor{blue}{B}\textcolor{teal}{D}$_3$   & \parbox{3.5cm}{\textcolor{purple}{A}=Sr,Ba,La \textcolor{blue}{B}=Co,Rh,Ir,Ni,Pt,Au,\\[-0.5\baselineskip] \textcolor{teal}{D}=Si,Ge,Sn}                              & BaPtSi$_3$                    & (101)-\textcolor{purple}{A}, (001)-Ba             & 1.62      & \parbox{3cm}{\vspace{.1\baselineskip}1068559,1068048,11879,\\[-0.5\baselineskip]30433,13123,1068447,2914,\\[-0.5\baselineskip]1070137,1069809,20910,\\[-0.5\baselineskip]1067925,22346,1070124,\\[-0.5\baselineskip]1069139,1070247,1069708} & 0         \\ \hline
123        &        &                                                                         & LiPdH                      & (100)-H-Li                    & 2.19      & 1018133                                                                                                                                                         & 0         \\ \cline{2-8} 
           &        &                                                                         & CaFeO$_2$                     & (100)-Ca-O                    & 2.31      & 19842                                                                                                                                                           & 0     \\ \hline
139        & \textcolor{purple}{A}\textcolor{blue}{B}$_2$    & \textcolor{purple}{A}=Rb,Ca,Sr \textcolor{blue}{B}=N,O                                                        & SrN$_2$                       & \parbox{2cm}{(101)-\textcolor{purple}{A}, (111)-N,\\[-0.5\baselineskip] (110)-\textcolor{purple}{A}+\textcolor{blue}{B}}   & 1.61      & \parbox{3cm}{12105,10564,2697,\\[-0.5\baselineskip]1009657}                                                                                                                      & 0-2.9    \\ \cline{2-8} 
           & \textcolor{purple}{A}\textcolor{blue}{B}$_2$\textcolor{teal}{D}$_2$  & \parbox{3.5cm}{\textcolor{purple}{A}=Cs,Ba,La,Rh,\\[-0.5\baselineskip] \textcolor{blue}{B}=Mg,Co,Rh,Ni,Pd,Pt,Cu,Ag,Zn,Al,\\[-0.5\baselineskip] \textcolor{teal}{D}=Si,Ge,Sn,P,As,Sb,S,Se} & CsCo$_2$Se$_2$                   & \parbox{2cm}{(101)-\textcolor{purple}{A}, (101)-Li+N,\\[-0.5\baselineskip] (001)-Ba} & 1.67      & \parbox{3cm}{\vspace{.1\baselineskip}9473,31059,8583,21057,\\[-0.5\baselineskip]560663,1070267,7882,\\[-0.5\baselineskip]6963,7875,12863,9247,\\[-0.5\baselineskip]9610,571343,6962}                                          & 0-3.8    \\ \cline{2-8} 
           & \textcolor{purple}{A}\textcolor{blue}{B}$_2$\textcolor{teal}{D}$_2$  & \parbox{3.5cm}{\textcolor{purple}{A}=Zn,Pd,Bi, \textcolor{blue}{B}=Ca,Ba,Sr,Li,Na,La,\\[-0.5\baselineskip] \textcolor{teal}{D}=N,H,O}                              & ZnBa$_2$N$_2$                    & \parbox{2cm}{(001)-\textcolor{blue}{B}+\textcolor{teal}{D}, (101)-Sr,\\[-0.5\baselineskip] (101)-O}  & 1.42      & \parbox{3cm}{8818,9307,9306,\\[-0.5\baselineskip]644389,23954,1070601}                                                                                                              & 0-0.9    \\ \cline{2-8} 
           & \textcolor{purple}{A}\textcolor{blue}{B}$_4$    & \textcolor{purple}{A}=Rb,Ba, \textcolor{blue}{B}=Al,Ga,In                                                     & BaAl$_4$                      & (001)-Rb, (101)-Ba            & 1.96      & 21477,22687,1903,335                                                                                                                                  & 0         \\ \hline
160        & \textcolor{purple}{A}IO$_3$   & \textcolor{purple}{A}=K,Rb                                                                  & RbIO$_3$                      & (101)-\textcolor{purple}{A}+O, (001)-I            & 1.35      & 27193,552729                                                                                                                                                 & 3.0-3.2   \\ \hline
164        &        &                                                                         & BaSi$_2$                      & (100)-Ba                      & 1.77      & 7655                                                                                                                                                            & 0         \\ \cline{2-8} 
           & \textcolor{purple}{A}\textcolor{blue}{B}$_2$C$_2$  & \textcolor{purple}{A}=Pd,Pt, \textcolor{blue}{B}=K,Rb,Cs                                                      & PdRb$_2$C$_2$                    & (100)-\textcolor{blue}{B}, (001)-\textcolor{blue}{B}              & 0.93      & \parbox{3cm}{976876,505825,505824,\\[-0.5\baselineskip]10918\vspace{.1\baselineskip}}                                                                                                & 1.7-2.1 \\ \cline{2-8} 
           & \textcolor{purple}{A}\textcolor{blue}{B}$_2$\textcolor{teal}{D}$_2$  & \textcolor{purple}{A}=Mg,Zr,Mn, \textcolor{blue}{B}=Li,Be, \textcolor{teal}{D}=N,O                                              & ZrLi$_2$N$_2$                    & (001)-\textcolor{blue}{B}                       & 1.82      & 19279,3216,11917                                                                                                                                          & 1.6-4.1 \\ \hline
166        & \textcolor{purple}{A}\textcolor{blue}{B}\textcolor{teal}{D}$_2$   & \parbox{3.5cm}{\textcolor{purple}{A}=K,Rb,Cs,\\[-0.5\baselineskip] \textcolor{blue}{B}=Bi,Y,La,Pr,Sm,Gd,Tb,Ho,Er,Lu,Cr,Tl,\\[-0.5\baselineskip] \textcolor{teal}{D}=O,S,Se,Te}            & RbBiS$_2$ & (101)-\textcolor{purple}{A}, (101)-\textcolor{purple}{A}+\textcolor{teal}{D}            & 1.41      & \parbox{3cm}{\vspace{.1\baselineskip}30041,8175,16763,9085,\\[-0.5\baselineskip]11739,561586,4026,\\[-0.5\baselineskip]9362,10780,10782,\\[-0.5\baselineskip]9364,9367,10783,\\[-0.5\baselineskip]7045,561619,9082}                          & 0.1-2.6 \\ \hline
186        &        &                                                                         & LiI                        & (101)-Li, (001)-I             & 1.42      & 570935                                                                                                                                                          & 4.4      \\ \hline
187        & Ba\textcolor{purple}{A}\textcolor{blue}{B}   & \textcolor{purple}{A}=Ge,As,Sb, \textcolor{blue}{B}=Pd,Pt,Al                                                  & BaSbPt                     & (100)-Ba                      & 1.81      & 8606,13272,9744                                                                                                                                           & 0         \\ \hline
191        &        &                                                                         & BaSi$_2$                      & (111)-Ba                      & 1.89      & 7701                                                                                                                                                            & 0         \\ \hline
194        & \textcolor{purple}{A}\textcolor{blue}{B}     & \textcolor{purple}{A}=Li,Ba, \textcolor{blue}{B}=Pd,Pt,B                                                      & BaPt                       & (100)-\textcolor{purple}{A}                       & 1.72      & 1064367,31498,1001835                                                                                                                                     & 0         \\ \hline
221        & \textcolor{purple}{A}\textcolor{blue}{B}     & \textcolor{purple}{A}=Rb,Cs,Cl, \textcolor{blue}{B}=Au,Tl,Br                                                  & RbAu                       & (111)-\textcolor{purple}{A}                       & 1.42      & 2667,30373,22906,23167                                                                                                                                 & 0.7-4.8 \\ \cline{2-8} 
           & \textcolor{purple}{A}\textcolor{blue}{B}\textcolor{teal}{D}$_3$   & \parbox{3.5cm}{\textcolor{purple}{A}=K,Rb,Cs,Sr,Ba,\\[-0.5\baselineskip] \textcolor{blue}{B}=Mg,Ca,Zr,Hf,V,Cr,Mn,Sn,\\[-0.5\baselineskip] \textcolor{teal}{D}=H,O,F,Cl,Br}                & BaVO$_3$                      & (110)-\textcolor{blue}{B}+\textcolor{teal}{D}, (100)-\textcolor{purple}{A}+\textcolor{teal}{D}          & 1.16      & \parbox{3cm}{\vspace{.1\baselineskip}27214,1070375,600089,\\[-0.5\baselineskip]23949,644203,3323,\\[-0.5\baselineskip]558749,23737,3834,\\[-0.5\baselineskip]1017465,20029,4551}                                           & 0-3.8    \\ \hline
225        & \textcolor{purple}{A}\textcolor{blue}{B}     & \textcolor{purple}{A}=Li,K,Rb,La,Pr, \textcolor{blue}{B}=H,S,Se,Cl                                            & KH                         & \parbox{2.1cm}{(110)-\textcolor{purple}{A}+\textcolor{blue}{B}, (100)-\textcolor{purple}{A}+\textcolor{blue}{B},\\[-0.5\baselineskip] (100)-\textcolor{blue}{B}} & 1.76      & \parbox{3cm}{24721,2350,24084,2495,\\[-0.5\baselineskip]23703,1161,22905}                                                                                                        & 0-6.4    \\ \cline{2-8} 
           & \textcolor{purple}{A}$_2$\textcolor{blue}{B}    & \textcolor{purple}{A}=Li,Na,K,Rb, \textcolor{blue}{B}=Pd,O,S,Se,Te                                            & Rb$_2$S                       & (111)-\textcolor{purple}{A}                       & 1.19      & \parbox{3cm}{\vspace{.1\baselineskip}2784,1394,11327,971,\\[-0.5\baselineskip]2352,1022,8041,2530,\\[-0.5\baselineskip]8426,1266,1747,648,\\[-0.5\baselineskip]1960,2286,1062711}                                          & 1.1-5.0  \\ \cline{2-8} 
           & \textcolor{purple}{A}H$_2$    & \textcolor{purple}{A}=La,Pr                                                                 & LaH$_2$                       & (110)-\textcolor{purple}{A}+H                     & 1.80       & 24153,24095                                                                                                                                                  & 0         \\ \cline{2-8} 
           & \textcolor{purple}{A}H$_3$    & \textcolor{purple}{A}=La,Ce                                                                 & LaH$_3$                       & (110)-H                       & 2.07      & 1018144,1008376                                                                                                                                              & 0         \\ \hline
\end{tabular}
\caption{\linespread{1}\selectfont{}Surfaces with ultra-low work functions from previous version of the database. Similar materials are grouped together and the composition with the lowest work function is listed for each group. Work functions and bandgaps (PBE, from Materials Project) are in eV.}
\label{tab:lowWF}
\end{table}

\begin{table}
\tiny
\begin{tabular}{llllllll}
\hline
\textbf{sg} & \multicolumn{2}{l}{\textbf{Material Family}}                                       & \parbox{1cm}{\textbf{Comp. w/}\\[-0.5\baselineskip] \textbf{highest} \boldmath$\phi$} & \parbox{2cm}{\textbf{Miller planes -}\\[-0.5\baselineskip] \textbf{Termination}  } & \boldmath$\phi$ & \textbf{mpid (mp-...)}                                                                                                                                                               & \boldmath$E_\mathrm{g}$   \\ \hline
8          &                        &                                                        & KCN                        & (110)-C, (001)-C            & 8.36      & 20134                                                                                                                    & 5.1     \\ \hline
12         &                        &                                                        & LiCuO$_2$                     & ({\sansmath $1\bar{1}0$})-O                    & 8.16      & 9158                                                                                                                     & 0.4     \\ \cline{2-8} 
           &                        &                                                        & Na$_2$CN$_2$                     & (001)-N                     & 8.96      & 541989                                                                                                                   & 3       \\ \hline
38         & \textcolor{purple}{A}\textcolor{blue}{B}O$_3$ & \textcolor{purple}{A}=K,Ba,Zr, \textcolor{blue}{B}=Ti,Nb,Pb                                  & ZrPbO$_3$                     & (010)-O, (110)-O            & 9.44      & 20337,5777,5246                                                                                                    & 2.1-3.3 \\ \hline
39         &                        &                                                        & TlF                        & (010)-F                     & 9.11      & 558134                                                                                                                   & 3       \\ \hline
44         &  &                                                        & NaO$_3$                       & (010)-O, (001)-O            & 8.80      & 22464                                                                                                                    & 0.6     \\ \hline
63         &                        &                                                        & TlCl                       & (010)-Cl                    & 8.07      & 571079                                                                                                                   & 2.7     \\ \hline
66         &                        &                                                        & AgO                        & (001)-O                     & 8.11      & 499                                                                                                                      & 0       \\ \hline
99         &                        &                                                        & KNbO$_3$                      & (001)-O                     & 9.47      & 4342                                                                                                                     & 1.6     \\ \hline
139        & \textcolor{purple}{A}\textcolor{blue}{B}$_2$                    & \textcolor{purple}{A}=K,Rb,Cs,Ca,Sr, \textcolor{blue}{B}=N,O                                 & CaO$_2$                       & (001)-\textcolor{blue}{B}                     & 9.68      & \parbox{3cm}{1441,12105,1866,\\[-0.5\baselineskip]1009657,634859,2697}                                                                       & 0-2.9   \\ \cline{2-8} 
           &                        &                                                        & Bi$_2$SeO$_2$                    & (101)-O                     & 8.34      & 552098                                                                                                                   & 0.4     \\ \cline{2-8} 
        & \textcolor{purple}{A}F$_4$                    & \textcolor{purple}{A}=Sn,Pb                                                & SnF$_4$                       & (100)-F, (101)-F, (110)-F   & 10.77     & 341,2706                                                                                                       & 2.0-3.2 \\ \hline
155        &                        &                                                        & ScF$_3$                       & (110)-F                     & 8.15      & 559092                                                                                                                   & 6.1     \\ \hline
160        & \textcolor{purple}{A}TlO$_3$                  & \textcolor{purple}{A}=Br,I                                                 & ITlO$_3$                      & (101)-O                     & 8.81      & 29798,22981                                                                                                           & 3.1-3.7 \\ \cline{2-8} 
           & \textcolor{purple}{A}\textcolor{blue}{B}O$_3$                   & \textcolor{purple}{A}=K,Rb, \textcolor{blue}{B}=Br,I                                         & RbIO$_3$                      & (001)-O, (101)-O            & 8.81      & 22958,27193,552729                                                                                        & 3.0-4.1 \\ \cline{2-8} 
           &                        &                                                        & BaCO$_3$                      & (101)-O                     & 8.71      & 4559                                                                                                                     & 4.4     \\ \cline{2-8} 
           &                        &                                                        & BaTiO$_3$                     & (101)-O, (001)-O            & 9.45      & 5020                                                                                                                     & 2.6     \\ \hline
164        & \textcolor{purple}{A}$_2$\textcolor{blue}{B}\textcolor{teal}{D}$_2$                  & \textcolor{purple}{A}=Mg,La, \textcolor{blue}{B}=Br,S,Se, \textcolor{teal}{D}=N,O                              & La$_2$SO$_2$                     & (001)-\textcolor{teal}{D}                     & 9.18      & 11917,7233,4511                                                                                                    & 3.1-4.1 \\ \cline{2-8} 
           & \textcolor{purple}{A}$_2$O$_3$                   & \textcolor{purple}{A}=La,Bi                                                & Bi$_2$O$_3$                      & (001)-O                     & 9.25      & 1017552,1968                                                                                                          & 1.4-3.9 \\ \hline
166        & \textcolor{purple}{A}\textcolor{blue}{B}\textcolor{teal}{D}$_2$                   & \parbox{3.5cm}{\textcolor{purple}{A}=Na,K,Rb,Sr,\\[-0.5\baselineskip] \textcolor{blue}{B}=Sc,Y,La,Zr,Nb,Ta,Mo,Rh,Hg,Al,Tl,\\[-0.5\baselineskip] \textcolor{teal}{D}=N,O} & NaScO$_2$                     & (001)-\textcolor{teal}{D}                     & 9.62      & \parbox{3cm}{\vspace{.1\baselineskip}9382,5475,3056,8188,\\[-0.5\baselineskip]7958,7017,7748,8145,\\[-0.5\baselineskip]7914,8409,978857,\\[-0.5\baselineskip]8830,578610} & 0.3-4.8 \\ \hline
186        & \textcolor{purple}{A}\textcolor{blue}{B}                     & \textcolor{purple}{A}=Mg,Al,Ga,In,Si, \textcolor{blue}{B}=C,N,O                              & AlN                        & (001)-\textcolor{blue}{B}, (101)-\textcolor{blue}{B}            & 9.23      & 22205,7140,804,549706,661                                                                                    & 0.5-4.1 \\ \hline
216        & \textcolor{purple}{A}\textcolor{blue}{B}                     & \textcolor{purple}{A}=Zn,Si, \textcolor{blue}{B}=C,O                                         & ZnO                        & (111)-\textcolor{blue}{B}                     & 9.80      & 8062,1986                                                                                                             & 0.6-1.6 \\ \hline
221        & \textcolor{purple}{A}\textcolor{blue}{B}F$_3$                   & \textcolor{purple}{A}=K,Rb,Cs,Ba, \textcolor{blue}{B}=Li,Mg,Ca,Cd                            & BaLiF$_3$                     & (110)-F, (110)-\textcolor{blue}{B}+F          & 9.05      & \parbox{3cm}{\vspace{.1\baselineskip}8399,7104,3654,6951,\\[-0.5\baselineskip]8402,3448,10175,4950,\\[-0.5\baselineskip]8401,10250}                                  & 3.6-7.2 \\ \cline{2-8} 
           &                        &                                                        & CsCdCl$_3$                    & (110)-Cl                    & 8.02      & 568544                                                                                                                   & 1.9     \\ \cline{2-8} 
           & \textcolor{purple}{A}\textcolor{blue}{B}$_3$                    & \textcolor{purple}{A}=Sc,W,Re,Al, \textcolor{blue}{B}=O,F                                    & AlF$_3$                       & (111)-\textcolor{blue}{B}, (110)-\textcolor{blue}{B}, (100)-O   & 11.04     & \parbox{3cm}{19390,190,10694,8039}                                                         & 0-7.7   \\ \cline{2-8} 
           &                        &                                                        & CsCl                       & (100)-Cl                    & 8.13      & 22865                                                                                                                    & 5.5     \\ \hline 
225        & \textcolor{purple}{A}\textcolor{blue}{B}                     & \textcolor{purple}{A}=Li,Na,K,Rb,Cs,Ca,Cd, \textcolor{blue}{B}=O,F,Cl                        & NaF                        & (111)-\textcolor{blue}{B}                     & 10.40     & \parbox{3cm}{23295,573697,1784,2605,\\[-0.5\baselineskip]1132,22905,682,463}                                                           & 0-6.4        \\ \cline{2-8}
           & \textcolor{purple}{A}\textcolor{blue}{B}$_2$                    & \textcolor{purple}{A}=Sr,Ba,Cd,Hg,Pb,Bi, \textcolor{blue}{B}=O,F,Cl                          & CdF$_2$                       & (111)-\textcolor{blue}{B}, (100)-\textcolor{blue}{B}            & 10.80     & \parbox{3cm}{568662,23209,20158,315,\\[-0.5\baselineskip]32548,241,8177            } & 0-5.6     \\  \hline
\end{tabular}
\caption{\linespread{1}\selectfont{}Surfaces with ultra-high work functions from previous version of the database. Similar materials are grouped together and the composition with the lowest work function is listed for each group. Work functions and bandgaps (PBE, from Materials Project) are in eV.}
\label{tab:highWF}
\end{table}